\documentclass[sigconf]{acmart}

\setcopyright{none}
\acmYear{2023}
\acmConference[]{}{}{}

\usepackage{tikz,graphicx,xcolor}
\usepackage{gensymb}
\usepackage{textcomp}
\usepackage{float,dblfloatfix}
\usepackage{verbatim,array,enumitem}
\usepackage{algorithm,algpseudocode,breqn,amsmath}
\usepackage{adjustbox,diagbox,threeparttable,tablefootnote,colortbl}
\usepackage{booktabs,tabularx,tabu,multirow,multicol,hhline}
\usepackage{caption,subcaption}
\PassOptionsToPackage{hyphens}{url} 
\usepackage{hyperref}
\usepackage[capitalise,noabbrev]{cleveref} 
\crefname{appsec}{Appendix}{Appendices} 
\usepackage[misc]{ifsym}
\usepackage[utf8]{inputenc}

\begin{document}

\title{Exploiting Out-of-band Motion Sensor Data to De-anonymize Virtual Reality Users}

\author{Mohd Sabra}
\email{mohd.sabra@utsa.edu}
\affiliation{%
  \institution{University of Texas at San Antonio}
}
\author{Nisha Vinayaga Sureshkanth}
\email{vsnisha@ieee.org}
\affiliation{%
  \institution{University of Texas at San Antonio}
}
\author{Ari Sharma}
\email{arisharma2017@gmail.com}
\affiliation{%
  \institution{Liberal Arts and Science Academy}
}
\author{Anindya Maiti}
\email{am@ou.edu}
\affiliation{%
  \institution{University of Oklahoma}
}
\author{Murtuza Jadliwala}
\email{murtuza.jadliwala@utsa.edu}
\affiliation{%
  \institution{University of Texas at San Antonio}
}


\begin{abstract}

Virtual Reality (VR) is an exciting new consumer technology which offers an immersive audio-visual experience to users through which they can navigate and interact with a digitally represented 3D space (i.e., a \emph{virtual world}) using a headset device. By (visually) transporting users from the real or physical world to exciting and realistic virtual spaces, VR systems can enable true-to-life and more interactive versions of traditional applications such as gaming, remote conferencing, social networking and virtual tourism. However, as with any new consumer technology, VR applications also present significant user-privacy challenges. This paper studies a new type of privacy attack targeting VR users by connecting their activities visible in the virtual world (enabled by some VR application/service) to their physical state sensed in the real world. Specifically, this paper analyzes the feasibility of carrying out a \emph{de-anonymization} or \emph{identification} attack on VR users by \emph{correlating} visually observed movements of users' avatars in the virtual world with some auxiliary data (e.g., motion sensor data from mobile/wearable devices held by users) representing their context/state in the physical world. To enable this attack, this paper proposes a novel framework which first employs a learning-based activity classification approach to translate the disparate visual movement data and motion sensor data into an \emph{activity-vector} to ease comparison, followed by a filtering and identity ranking phase outputting an ordered list of potential identities corresponding to the target visual movement data. Extensive empirical evaluation of the proposed framework, under a comprehensive set of experimental settings, demonstrates the feasibility of such a de-anonymization attack.

\end{abstract}

\maketitle

\section{Introduction}
\label{sec:introduction}

Virtual Reality (VR) is changing the paradigms of human-computer interaction, and has become a ubiquitous consumer technology \cite{metaverse,spatial,meetinvr,mozillahubs,sketchfab,googleblocks,vrchat}. Most prevalent VR systems today (e.g., Meta Quest, HP Reverb and Sony PlayStation VR), offer an immersive audio-visual experience where users can navigate around a digitally represented 3D space (i.e., a \emph{virtual world}) using a VR headset. In most VR systems, users would navigate and interact with this virtual world using on-body (often, handheld) controllers that can track users' body movements in the real world and execute analogous movements in the virtual world. Reactions from users' navigation actions and interactions with the virtual world are relayed back to the user by means of video, audio, and haptic (e.g., vibrations) signals perceptible to the user through his/her VR device/headset.

VR systems enable several novel applications that were previously not possible using traditional desktop and mobile devices, such as immersive gaming \cite{oculusstore}, remote conferencing \cite{meetinvr,mozillahubs}, virtual tourism \cite{VRTour,VRTravel}, social networking \cite{servreality,metaHorizon}, visualizing 3D models \cite{sketchfab,googleblocks}, and large open-world spaces \cite{spatial,metaverse}. Unfortunately, at the same time new privacy and security challenges have also emerged in the VR space. For example, password inference from finger movements (using motion sensors) when typing a password in the virtual world can become a security problem if the same password is reused by the user in real world \cite{das2014tangled}. Some VR headsets also include eye-tracking, which can become an additional channel for inference of private data. For instance, it has been shown in recent research efforts that eye tracking or gaze data could be potentially misused to infer a user's personal information and traits such as gender, age, ethnicity, body weight, personality traits, drug consumption habits, emotional state, skills and abilities, fears, interests, and sexual preferences \cite{kroger2019does}. Personal gait and movement data collected from a VR headset can also be used in conjunction with Deepfake videos to create highly authentic looking fake videos \cite{westerlund2019emergence}, which can be further used to damage personal reputation \cite{dhsdeepfake,mrdeepfakes}, conduct social engineering attacks \cite{wojewidka2020deepfake,tripwiredfse}, and spread misinformation (fake news) \cite{diakopoulos2021anticipating,karnouskos2020artificial}.

VR systems and applications can transport users from the real world to a virtual world, wherein the two worlds are seemingly disconnected from each other. \emph{In this work, we study the potential of a new type of privacy attack targeting VR users by connecting their activities visible in the virtual world (enabled by some VR application/service) to their physical state sensed in the real world.} More specifically, we analyze the potential of carrying out a %
\emph{de-anonymization} or \emph{identification} attack on VR users by \emph{correlating} visually observed movements of users' anonymous humanoid avatars in the virtual world with auxiliary data representing their context/state in the physical world. For such auxiliary information, we specifically focus on data available from motion sensors on-board mobile and wearable devices (e.g., smartphones and smartwatches) that users may be carrying on them while navigating or interacting with virtual worlds in VR applications. Our attack is motivated by the fact that motion sensors on-board modern mobile devices, such as, accelerometers and gyroscopes, are considered to be \emph{zero-permission}, i.e., any on-device application can record/sample\footnote{Android 12+ requires the \texttt{HIGH_SAMPLING_RATE_SENSORS} permission to sample motion sensors beyond 200 $Hz$. However, our correlation framework can operate with a sampling frequency much lower than 200 $Hz$.} data from these sensors without requiring explicit user-permissions. This, consequently, enables easy misuse of such motion data by any on-device app, something which has been extensively documented in the security research literature \cite{cai2011touchlogger,owusu2012accessory,maiti2016smartwatch,liu2015good,han2012accomplice,mosenia2017pinme,michalevsky2014gyrophone,han2017pitchln,song2016my,davarci2017age,singh2019side}. 
Moreover, we hypothesize that fine-grained user movement captured by on-body motion sensors (such as those on a user's wrist in the form of a smartwatch and in a user's pocket in the form of a smartphone) are strongly correlated with the visual motions observed in the user's humanoid avatar in the virtual environment of a VR app, and thus can potential be used to de-anonymize users in virtual spaces.

We consider an attack scenario where the adversary is trying to de-anonymize a target user in the virtual world by visually tracking the motion of the user's avatar and then correlating it with labeled motion data streams belonging to a (large) set of users, which also includes motion data from the target user. We refer to this set of labeled motion data streams (belonging to a large set of users) as the target user's \emph{anonymity set}. Such an attack scenario could arise in many popular VR services such as Metaverse \cite{metaverse}, VRChat \cite{vrchat} and MeetinVR \cite{meetinvr} which enable a group of users to organize events, get-togethers and games in some virtual environment. A mobile (smartphone) app of an adversarial service provider can be used to stealthily record motion data of every user in the group for a particular event, while video data corresponding to a target user (in the group) can be captured by the adversary directly from the virtual environment/world, say, by participating or entering the same virtual environment/world as the target user. We propose a novel correlation framework to carry out the de-anonymization in such VR services, and comprehensively evaluate parameters such as effect of different actions/movements and the effect of using having their mobile device in different bodily locations (such as different pockets). 

Our work advances research investigation of how data from the physical world can be used to compromise the privacy of users in the virtual worlds. We believe that this is the first research effort which investigates this issue. Identity protection is key to VR innovation as otherwise users will be hesitant to participate in the ecosystem \cite{adams2018ethics} in order to protect their privacy, reputation and security. To protect users from the potential de-anonymization attack, we also propose novel countermeasures that users can adopt while participating in VR applications. 
In summary, we make the following main contributions in this paper:
\begin{enumerate}[leftmargin=0.25in]
\item Contribution 1: A framework that transmutes both motion sensor data and avatar's visual movement data in to a comparable \emph{activity-vector}.
\item Contribution 2: A correlation model that filters mismatching activity-vectors, and ranks matching activity vectors from best to worst.
\item Contribution 3: Test data collection for real-world human participants, and a comprehensive empirical evaluation of the correlation framework under various settings.
\item Contribution 4: Improvements and optimizations of the correlation framework for a \emph{large-scale} attack.
\end{enumerate}

\section{Related Work}
\label{sec:related}

Research efforts in the literature related to our work can be categorized into those that focus on inferring private information by means of mobile device motion sensors which we discuss first, followed by those that propose new sensitive information inference vectors for VR systems and applications.

\noindent \textbf{Information leakage through mobile device motion sensors: }
Mobile and wearable device motion sensors such as accelerometers and gyroscopes have been heavily scrutinized in the research literature for their potential to be employed as a side-channel for leaking users' private information. For instance, motion sensor data on smartphones and smartwatches have been utilized to infer keystrokes and passwords \cite{sarkisyan2015wristsnoop,cai2011touchlogger,owusu2012accessory,maiti2016smartwatch,liu2015good,lu2018snoopy}, identify lock screen patterns \cite{xu2012taplogger}, deduce travel routes and location \cite{narain2016inferring,han2012accomplice,mosenia2017pinme}, infer speeches \cite{hodges2018reconstructing,michalevsky2014gyrophone,han2017pitchln}, infer handwritten text \cite{wijewickrama2019dewristified}, reconstruct 3D models from printer vibrations \cite{song2016my}, and estimate demographic information \cite{davarci2017age,singh2019side}. Application of on-body motion sensors onboard consumer mobile devices such as smartphones and smartwatches for user authentication \cite{wijewickrama2021write,xu2016walkie,li2020handwritten} is another closely related research space that has received significant attention in the literature. However, such biometric authentication systems require training data from individual users, which is not available in our adversarial setting.

\noindent \textbf{Information leakage in VR systems and applications:}
Albeit relatively new as a consumer technology, VR has garnered a host of privacy and security concerns. As mentioned earlier, attacks such as password inference from finger movements (using motion sensors) when typing a password in the virtual world can become a security problem if the same password is reused by the user in real world \cite{das2014tangled}. Some VR headsets include eye-tracking, which can reveal valuable personal information \cite{kroger2019does}. VR, when used in conjunction with Deepfakes \cite{westerlund2019emergence}, can also become a serious threat as an adversary can potentially utilize personal gait and movement data collected from a VR headset to create a very authentic-looking fake video. These type of attacks can be used to damage personal reputation \cite{dhsdeepfake,mrdeepfakes}, conduct social engineering attacks \cite{wojewidka2020deepfake,tripwiredfse}, and spread misinformation (fake news) \cite{diakopoulos2021anticipating,karnouskos2020artificial}.

Authentication in VR is a closely related research topic \cite{stephenson2022sok}, wherein authorized sensors on the VR headset or paired on-body controllers are used to authenticate individual users. However, in our attack we focus on out-of-band motion sensor data, which are not paired with the VR system. De-anonymization solely using movements observed in the virtual world is difficult, especially when the anonymity set size is large. In this work, we carry out de-anonymization of users of a VR platform by correlating visually observed movements of the user's avatar in the virtual world with out-of-band motion data available from users. 

Use of anonymous avatars and identity transformation inside a virtual reality experience \cite{maloney2020anonymity,freeman2020my,gupta2020investigating} is a significant factor contributing to the technology's popularity. Therefore, identity protection is key to VR innovation as otherwise users will be hesitant to participate in the ecosystem \cite{adams2018ethics} in order to protect their privacy, reputation and security. Previous works on de-anonymization of VR users utilized \emph{in-band} data (such as sensors on the VR systems and/or movement characteristics of virtual avatars) to infer users' identity \cite{miller2020personal,miller2022combining}, anthropometrics \cite{nair2022exploring}, environment \cite{nair2022exploring}, device information \cite{trimananda2022ovrseen,nair2022exploring}, and demographics \cite{nair2022exploring}. To the best of our knowledge, our proposed de-anonymization attack using out-of-band motion data has thus far not been analyzed or publicly presented. %
We also evaluate the scope of the proposed de-anonymization attack within a small set of users and at a larger scale. To protect users from the proposed attack, we also suggest countermeasures that users can adopt while immersing in a VR experience.
\section{Threat Model}
\label{sec:threatmodel}

\begin{figure}[t]
 \centering
 \includegraphics[width=0.99\linewidth]{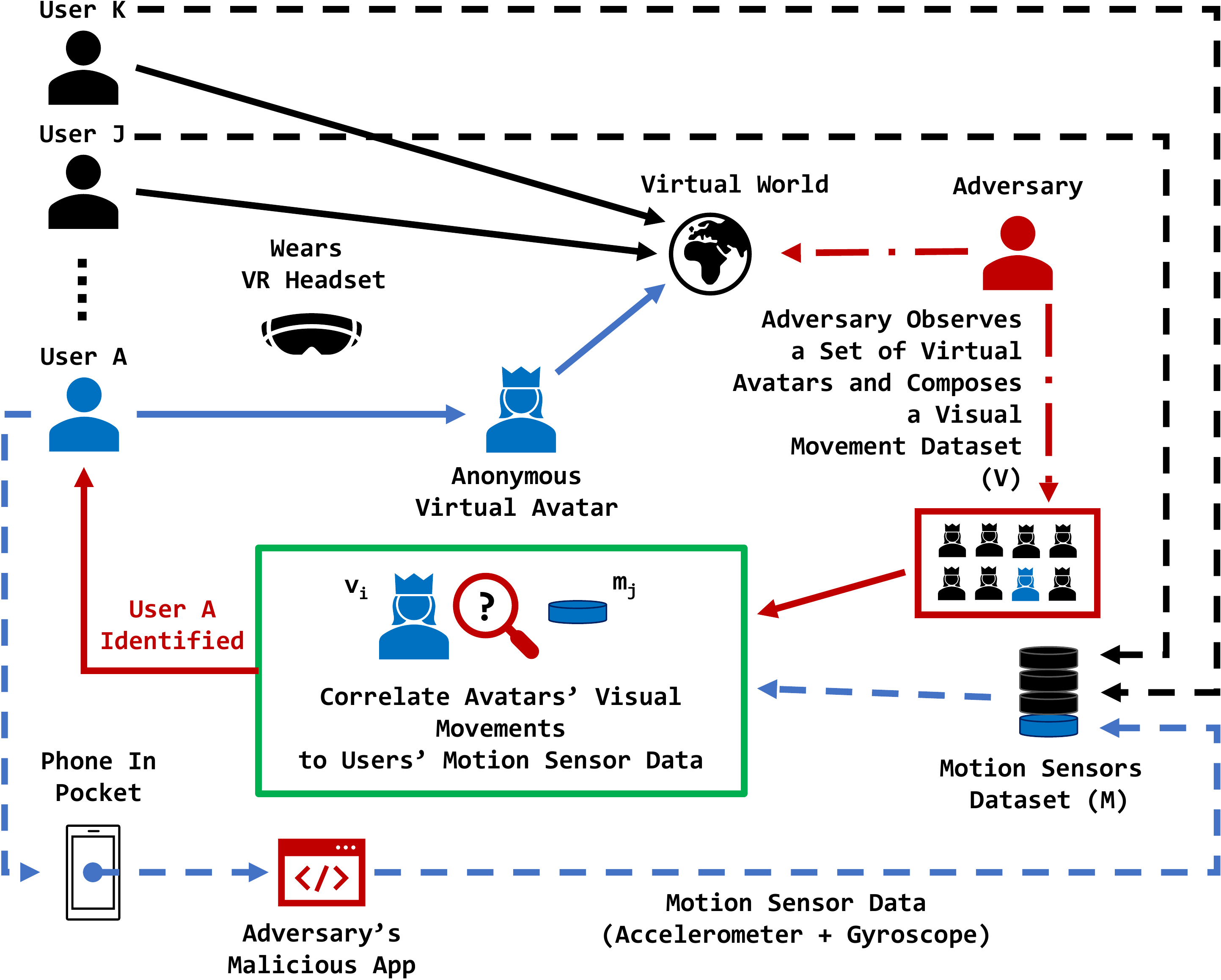}
 \caption{Threat model.}
 \label{fig:adversary}
 \vspace{-0.15in}
\end{figure}

We consider an adversary whose goal is to de-anonymize users of a VR ecosystem by correlating visual movements of anonymous virtual world avatars with out-of-band identifiable mobile/wearable motion sensor data from target users. 
The \emph{size} of the labeled motion dataset of users in the possession of the adversary, representing the \emph{anonymity set} of the target VR user or avatar, may vary between a \emph{large-scale} where the cardinality (of the dataset) may be very high, to a significantly smaller \emph{small-scale} (e.g., employees of a company or participants of an event). 
Similarly, the recordings of VR users or avatars will result in a visual movement dataset, which can also range between a \emph{global} scale where its cardinality may be very high, to a significantly smaller \emph{small-scale} such as avatars present within a (targeted) virtual room or playing a (targeted) virtual game.
As depicted in \cref{fig:adversary}, the goal of the adversary is to de-anonymize a target user (i.e., its avatar) in the VR space by matching an element in the labeled motion dataset to the element (corresponding to the target user or avatar) in the visual movement dataset by utilizing some efficient correlation mechanism, similar to the one we propose in \cref{sec:framework}. This adversarial goal can be easily extended to include de-anonymization of multiple VR users or avatars. 

In order to compile the visual movement dataset (denoted by $V=\{v_1,v_2,\ldots,v_p\}$, with cardinality $p$), the adversary has to join the virtual world, observe and record each avatar for a baseline duration of time within which a series of movements are likely observed. In case of the VR service provider being the adversary, this process can scale easily. %
In order to compile the labeled motion dataset (denoted by $M=\{m_1,m_2,\ldots,m_q\}$, with cardinality $q$), the adversary installs a malicious data collection app on the mobile/wearable devices of a targeted set of users, which records zero-permission motion (accelerometer and gyroscope) sensor data and reports it back to the adversary. As explained earlier, this targeted set of users can be at a small or large scale. Typically, this can be achieved by means of a trojan app that offers some utility to the users on the front-end (e.g., a game or a social networking service), while surreptitiously recording the motion data on the back-end. We also assume that both datasets ($V$ and $M$) contains timestamps which are fairly in sync with the standard global time. %

For popular apps/services that also offer a VR platform, for example, Meta, such an attack can potentially be scaled globally for both the motion and visual datasets. Nonetheless, such an attack is easier to be carried out at a \emph{small-scale}, implying that the malicious mobile/wearable app has to be popular within a small group of users and/or the VR ecosystem has to be popular within the group. When both $p$ and $q$ are large, the correlation process to de-anonymize all users grows to be computationally challenging for the adversary. In \cref{eval-scalability} we propose optimization techniques that can significantly reduce the computational complexity, and thus the average runtime, of the proposed correlation framework. Below we present two different scenarios representing our threat model.

\noindent
\textbf{Scenario 1.}
A large organization (such as Meta) that operates both a popular VR platform (such as Metaverse) and a popular mobile app (such as Facebook, WhatsApp, and Instagram) can collect both the visual movement dataset and motion sensor dataset for respective platforms. Users who do not want to be identified across both of these platforms are susceptible to the proposed de-anonymization attack, even when using anonymous identity and avatar on the VR platform. This scenario represents a \emph{large-scale} attack where the anonymity set is large.

\noindent
\textbf{Scenario 2.}
A criminal group uses VRChat \cite{vrchat} to anonymously meetup. An undercover police officer present in the meetups is able to record the visual movements of individual (anonymous) avatars of the criminal group. With the help of a popular smartphone app company (such as Google), the police is also able to collect identified motion sensor data from a list of known criminals and suspects. Thereafter, the proposed correlation framework can be used by the police to de-anonymize members of the group on VRChat. This scenario represents a \emph{small-scale} attack where the anonymity set is small.

\section{Correlation Framework}
\label{sec:framework}

\begin{figure}[t]
	\centering
	\includegraphics[width=0.99\linewidth]{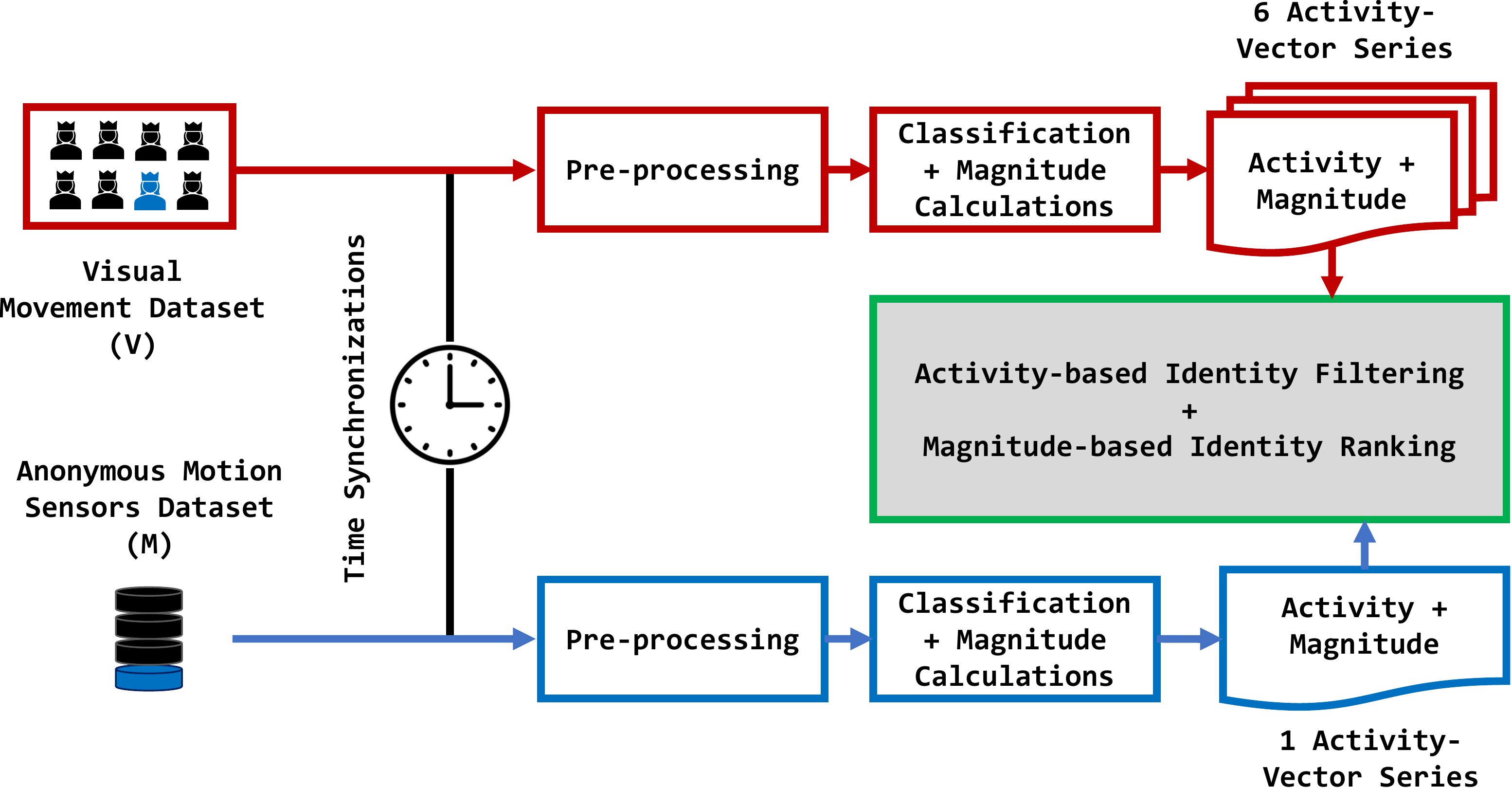}
	\caption{Overview of our correlation framework.}
	\label{fig:overview}
	\vspace{-0.15in}
\end{figure}

Our correlation framework (\cref{fig:overview}) is composed of two key components. The first component converts both the (out-of-band) motion sensor data and the visual movement data in to a comparable format, which we refer to as \emph{activity-vector series}. The activity-vector series enables us to directly compare and match elements from the two datasets ($V$ and $M$) using a matching heuristic. 
The second component in our framework ranks the closest matches across the elements of either dataset, in a fashion such that the high ranked matches are likely associated with the target user (identifiable from $M$).

\subsection{Activity-Vector Series}
\label{subsubsec-commondtatformat}

Our motivation behind defining a \emph{activity-vector series} stems from the fact that the two datasets (motion sensor data from the mobile/wearable device and visual movement dataset from the VR app) are not directly comparable to each other. The motion sensor data comprises of samples measuring linear acceleration and orientation changes of a user's body, whereas the visual movement data consists of a video wherein an anonymous avatar's movements are recorded as changes in pixels across its frames. 
Consequently, we define an \emph{activity-vector series} as a sequence of activities observed (classified by some machine learning or ML model as discussed later), combined with a pairwise sequence of ``magnitudes'' for each observed activity from each of the data sources (visual movements and motion sensor). Our magnitude quantification of an observed activity is approximate, but serves as a critical attribute in our correlation framework as detailed in \cref{subsubsec-magnitude}.

More precisely, our activity-vector series is composed of the following commonly observed activities: \emph{idle}, \emph{body rotation}, \emph{head rotation}, \emph{hand movements}, \emph{walking}, \emph{bending}, \emph{jumping}, and ``\emph{other}''. 
These were the common movements observed in over 2000 hours of activity data collected inside VRChat \cite{vrchat} by us (more details on data collection can be found in \cref{sec:setup}). 
These activity classifications combined with magnitude calculations form a vector-like representation where each observed activity has a corresponding magnitude information (similar to a vector which consists of direction and magnitude).
An activity-vector series from either sources can be depicted as follows:

\vspace{0.1in}
\noindent
\includegraphics[width=\linewidth]{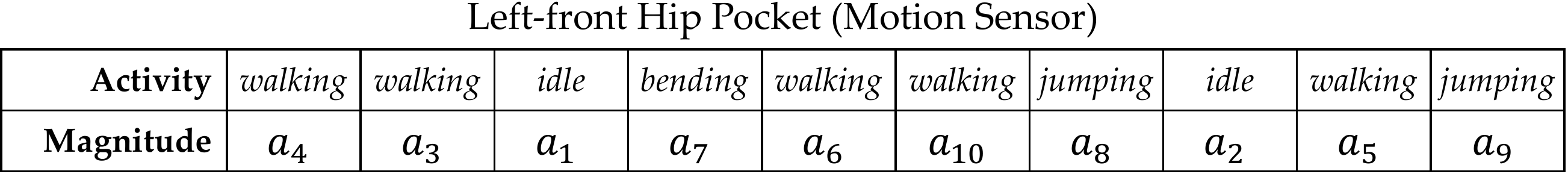}

\noindent
where $a_i\in\mathbb{R}^+$ is the positive real magnitude of an activity time window, such that $a_{1}>a_{2}>\ldots>a_{10}$.
In order to generate this activity-vector series, we next detail the steps taken to pre-process and utilize supervised machine learning models to classify the activities observed in individual sequences.

\subsection{Pre-Processing}
We first segment both the physical motion data (obtained from the mobile device motion sensors) and the visual movement data (obtained from the VR apps) into small time windows (of $w$ seconds each) and classify each window as one of the eight aforementioned actions. We empirically evaluate the effect of the size of $w$ on correlation accuracy in \cref{eval-parameterThresholds} and use the optimal value for rest of the evaluation.
For the visual movement data, we further separate individual user's avatar from the background, so as to better classify the movements of the avatar without any background noise. PaddleSeg \cite{paddleseg2019}, an open-source toolkit that applies image segmentation using different techniques, was used to segment out the individual avatars. More specifically, we used a pre-trained ORCNet model with HRNet backbone that was trained using the Cityscapes dataset \cite{ocrnet-hrnet-w48-paddle}.
For the motion sensor data, we apply a Savitzky-Golay filter \cite{press1990savitzky} to smooth the signals for noise reduction before classification. %

\subsection{Training Data Generation}
\label{training-data-gen}
\begin{figure}[t]
	\centering
	\includegraphics[width=0.99\linewidth]{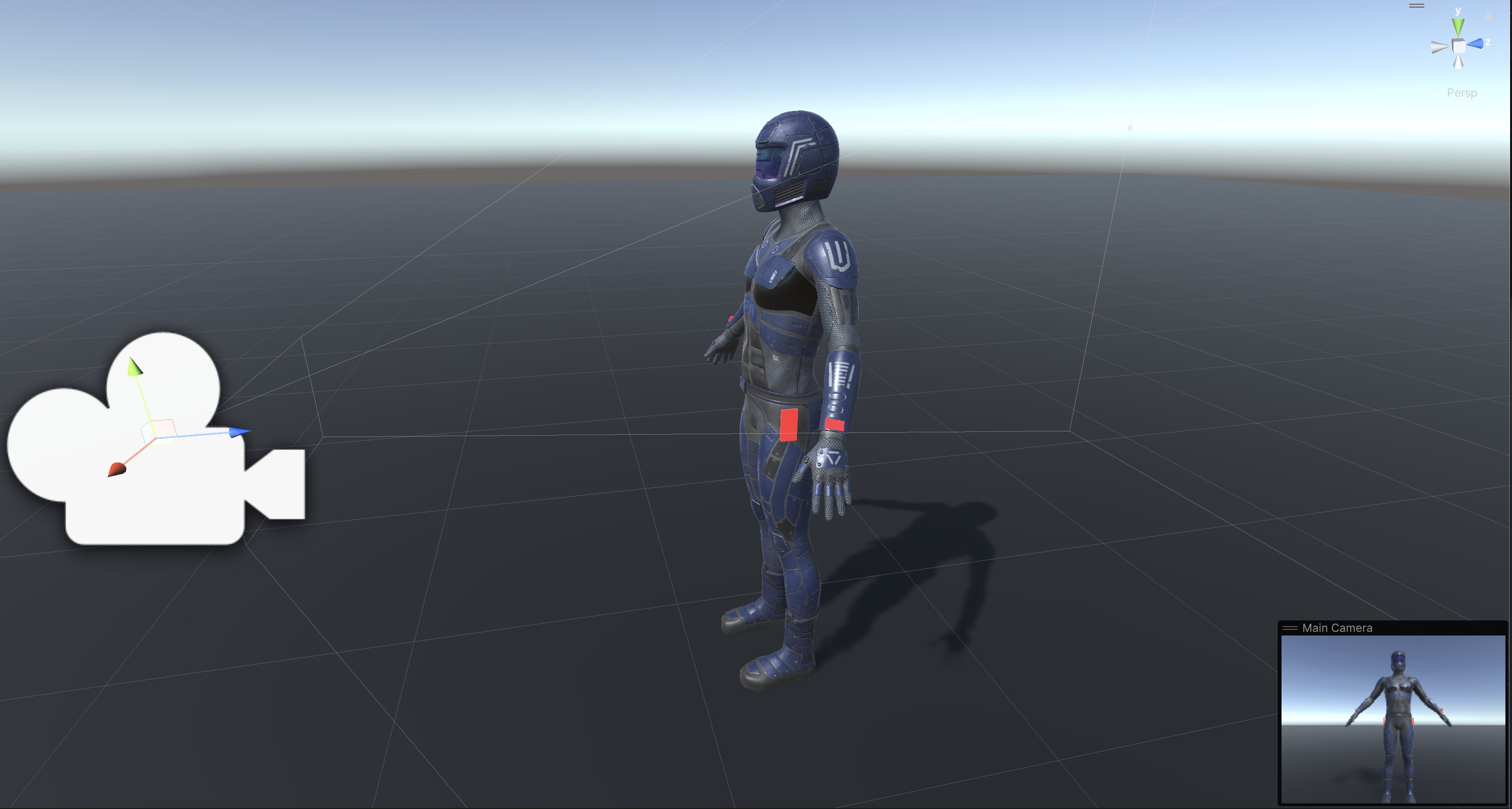}
	\caption{The training data generation setup inside Unity, depicting only one camera viewpoint and virtual motion sensors attached to the avatar (in red).}
	\label{fig:unitysetup}
	\vspace{-0.15in}
\end{figure}

In order to generalize and scale our activity classification for a \emph{large-scale} attack, %
we generate training data as an amalgamation of a well-known dataset in the literature and add synthetically generated variations to capture a wide range of bodily variances and anomalies (often caused by imperfections in the VR systems) all of which are otherwise impractical for collection from real human subjects. 
Specifically, we generate the training data of our visual movement classifier using the 3D game engine Unity \cite{unity} (\cref{fig:unitysetup}), utilizing the Carnegie Mellon University (CMU MoCap) \cite{cmu-mocap} dataset and synthetically generated variations of motions captured in the CMU MoCap dataset. 
The CMU MoCap dataset was created using a motion capture system where the subjects wore 41 markers and performed various activities. 
It is a well-known dataset for evaluation of activity recognition frameworks \cite{barnachon2014ongoing,mo2021keyframe,sigal2010humaneva}, %
and can be applied to reproduce avatar movements inside Unity using corresponding body keypoints (\cref{fig:keypoints}).

Our synthetically generated movement variations randomized the speed between $0.25\times$ and $2\times$ of CMU MoCap speeds, and rotation angle between $-10\degree$ and $+10\degree$ of CMU MoCap rotation angles. %
In addition to the CMU MoCap model avatar, we also train using another freely available avatar, namely the \emph{Futuristic soldier - Scifi character}\footnote{\url{https://assetstore.unity.com/packages/3d/characters/humanoids/sci-fi/futuristic-soldier-scifi-character-202085}}. 
As the video movement data is dependent on the viewpoint of the adversary, we also capture varying camera positions around the virtual avatar in Unity (\cref{fig:unitysetup}). 
Specifically, the camera position was randomized around the avatar (across all angles for which the avatar is visible), enabling a different visual perspective and thus improving our classifier training. The visual movements of the avatars were recorded using OBS Studio \cite{obsproject}.

Additionally, in Unity we attached a custom-made \emph{virtual} motion sensor to the avatar (\cref{fig:unitysetup}), which is able to capture acceleration and orientation changes of the avatar. This virtual motion sensor closely captures the kinematic forces experienced by the avatar in the same way a smartphone or smartwatch motion sensor on a real person would experience, and it allows us to collectively train a classifier for the motion sensor data alongside the visual movement classifier. The key advantages of using such a virtual sensor for training are the elimination of %
synchronization errors, and not requiring real human subject participants for data collection (except for the human subject participants who helped in the development of the CMU MoCap dataset).
Note that for our experimental evaluations (\cref{sec:eval}) with an adversarial standpoint, we compose a realistic \emph{test} dataset with the help of real human subject participants and also address synchronization errors between the motion sensor and visual movements data (\cref{eval-unsyncData}).

\subsection{Activity Classification}

We collectively utilize Apple's Core ML \footnote{\url{https://developer.apple.com/documentation/Core ML}} and Create ML\footnote{\url{https://developer.apple.com/documentation/createml}} libraries to generate two classification models (each trained separately), one for the video movement training data and another for the motion sensor training data. %
The Core ML model is already trained by Apple for generic action and activity classification, %
and can be further customized using transfer learning \cite{marques2020machine} using the training data generated in \cref{training-data-gen}. Prior research has already demonstrated the feasibility of such activity recognition using Core ML \cite{kumar2018combining}. Moreover, Apple's Vision framework\footnote{\url{https://developer.apple.com/documentation/vision}} is already pre-trained for keypoint detection on humans (\cref{fig:keypoints}), which can also be utilized with Core ML on humanoid avatars. %
\emph{Applying these trained classification models on test visual movement and motion sensor data split into $w$ second windows will result in a sequence of activities observed on the two data sources, which is one of the two sequences in the activity-vector series defined earlier.}

\subsection{Activity Magnitude}
\label{subsubsec-magnitude}

Intuitively, when the same classified activity is observed in both data sources (in a given time window), we can improve our identity correlation by \emph{ranking} smaller magnitude differences above larger magnitude difference. For example, if an anonymous avatar is observed to be walking fast in the virtual world (high magnitude), it is likely that their activity magnitude will also be high on the motion sensor data.
As mentioned earlier, our magnitude quantification of an observed activity is approximate. For the motion sensor, we calculate magnitude of each $w$ second activity window as the average magnitude of acceleration vectors in the motion sensor data. For the visual movement data, we utilize optical flow to compute the average acceleration of areas on the avatar's body where the motion sensor may be attached. Optical flow estimates the motion of objects between consecutive frames in a video, caused by the relative movement between the object and camera \cite{horn1981determining,decarlo1996integration}.

However, as some activities tend to generate disproportionate levels of motion in various parts of the body, it may result in different magnitudes of movements for the same activity. Furthermore, as the adversary may not have knowledge of the motion sensor's positioning for each user's data, the visually observed magnitude of movement experienced by an avatar's different body keypoints (\cref{fig:keypoints}) is another attribute that should be factored in to improve our correlation model.
We consider six usual body positions where the motion sensor is likely to be attached, such as a smartphone in pant pocket or a smartwatch on the wrist: 
left-front hip pocket,
right-front hip pocket,
left-back hip pocket,
right-back hip pocket,
left wrist, and
right wrist.
As a result, the activity-vector series calculated from the visual movement dataset will consists of six different magnitude sequences (for the same activity sequence) as follows:

\vspace{0.1in}
\noindent
\includegraphics[width=\linewidth]{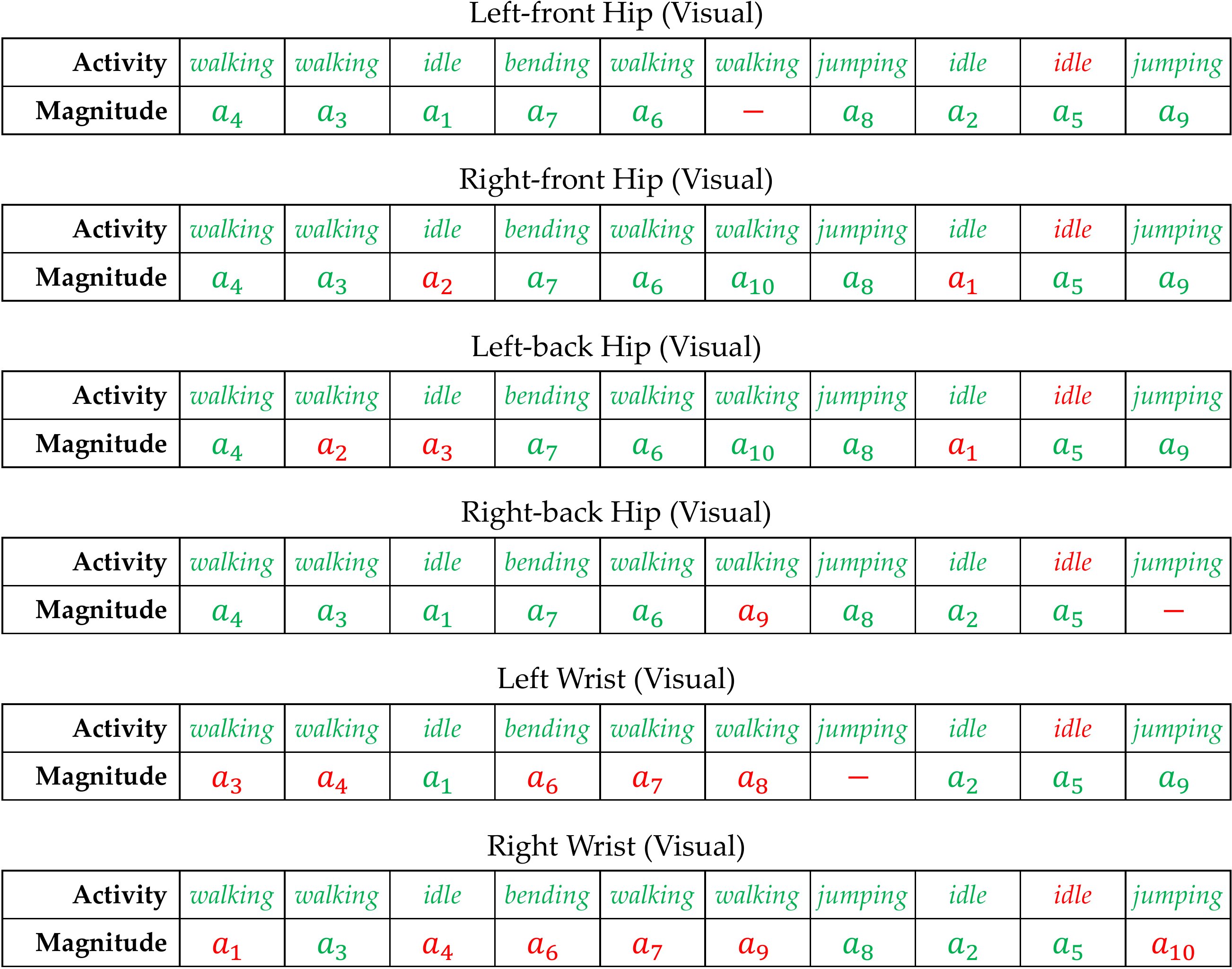}

\noindent
where {\color{red}``--''} implies unobservable position for optical flow calculations, all {\color{red}$a_i$} in red depict mismatched magnitude rank with the left-front hip pocket motion sensor activity-vector series shown in \cref{subsubsec-commondtatformat}, and all green {\color{green}$a_i$} imply matching magnitude rank. Moreover, there is an activity misclassification in this example at the ninth window, highlighted as {\color{red}$idle$} in red. All of these seven magnitude sequences (one from motion sensor data and six from visual movement data) are utilized in the correlation and identity ranking processes described next.

\subsection{Correlation and Identity Ranking}

The first intuitive assumption in our correlation framework is that the order of activities conducted by an user (and their avatar) will be unique when observed for a long enough duration. Intuitively, this observation duration can be shorter in a \emph{small-scale} attack where the anonymity set is smaller. In a \emph{large-scale} attack, the observation duration has to be longer because with a large anonymity set the occurrence of more than one anonymous user conducting the same sequence of activities within a short observation duration is more probable, thus creating confusion between them.
We use this first assumption to filter out unlikely matches from our identity ranking calculations, using the activity sequences in the activity-vector series.

Our second intuitive assumption is that varying activity magnitudes caused by disproportional levels of motion in various parts of the body can be utilized to identify closely correlated visual movement and motion sensor sequences. 
Accordingly, we utilize magnitude correlation rankings to rank known identities (from dataset $M$) such that users with motion sensor magnitude sequence closely matching to a visual movement magnitude sequence (best of the six visual positions) are ranked closer to $1$. 

\subsubsection{Activity-based Filtering}
As the activity classification is not perfect, we cannot reliably use the sequence of activities for correlation. Instead, we use a high degree of mismatch between sequences of activities (across visual movement and motion sensor data) to filter out identities whose motion sensor data are objectively different from an anonymous avatar being observed.
More specifically, we calculate the hamming distance between the motion sensor activity sequence and the visual movement activity sequence (which is the same for all six activity-vector series generated from the visual movement data). Thereafter, we eliminate pairs with distance threshold $>t$ from further magnitude-based identity rankings. We empirically evaluate threshold $t$ in \cref{eval-parameterThresholds} as part of our framework parameter optimizations.
For example, between the activity-vector series illustrated in \cref{subsubsec-commondtatformat} and \cref{subsubsec-magnitude}, this hamming distance is 1 (or 10\%) due to the activity mismatch in the ninth time window.

\subsubsection{Magnitude-based Ranking}
After filtering, we are left with identities whose motion sensor activity sequences closely matched at least one of the six visual movement activity sequences. %
We utilize Spearman's rank correlation coefficient \cite{zar1972significance} to correlate and rank potential identities based on magnitude sequences, which is computed as follows:
$$\rho = 1-{\frac {6 \sum d_i^2}{n(n^2 - 1)}}$$
where $n$ is the number of observations (of $w$ second windows) in the activity-vector series, and $d_i$ is the difference in the paired ranks of the two magnitudes (across the visual movement and motion sensor data sequences) at the $i_{th}$ time window. %

The higher the Spearman's rank correlation coefficient, the more likely the two sequences correlate to each other, and thus the corresponding identity from $M$ would be ranked closer to $1$ out of the $q$ (minus the identities that did not pass the activity-based filtering). 
As the adversary does not have positioning information of the motion sensor on the users' body, we compute Spearman's correlation coefficient for the six likely positioning of the motion sensors (\cref{subsubsec-magnitude}), and consider only the maximum for identity ranking. Between the examples shown in \cref{subsubsec-commondtatformat} and \cref{subsubsec-magnitude}, magnitude from the visual data sequence 
of the left-front hip will have the highest Spearman's correlation coefficient with the left-front hip pocket motion sensor magnitudes.

When activity-based filtering threshold $t$ is set very low (i.e., only tolerance for very minor or no mismatches in the activity sequences), it is also possible that all identities are eliminated from this magnitude-based raking, thus resulting in no identity ranking. The entire correlation procedure is digested in \cref{alg:keystroke-detection}.

\begin{figure*}[b]
	\centering	
	\begin{subfigure}[b]{0.99\textwidth}
		\centering
		\begin{subfigure}[b]{0.19\textwidth}
		\centering
		\includegraphics[width=0.99\linewidth]{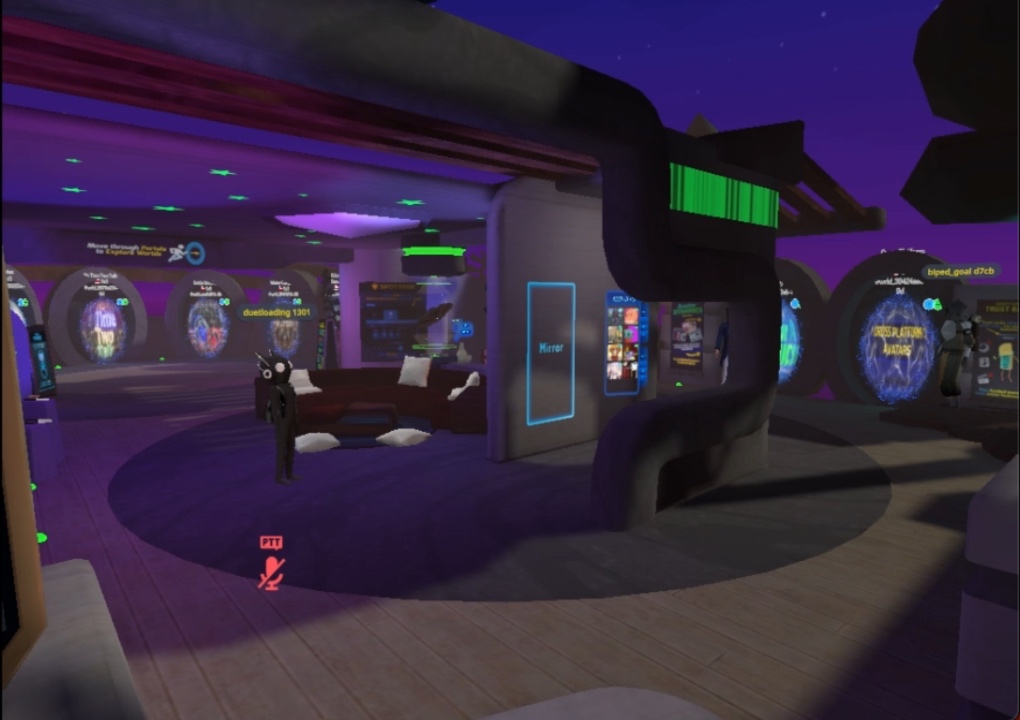}
		\caption{HC1}
	\end{subfigure}
		\begin{subfigure}[b]{0.19\textwidth}
		\centering
		\includegraphics[width=0.99\linewidth]{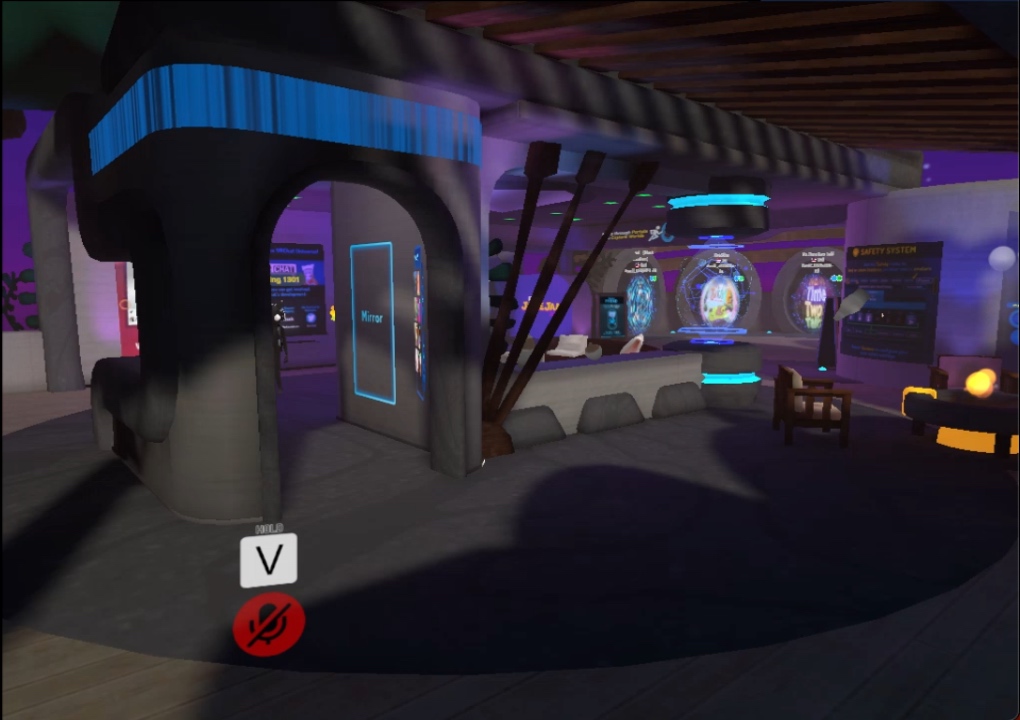}
		\caption{HC2}
	\end{subfigure}	\begin{subfigure}[b]{0.19\textwidth}
	\centering
	\includegraphics[width=0.99\linewidth]{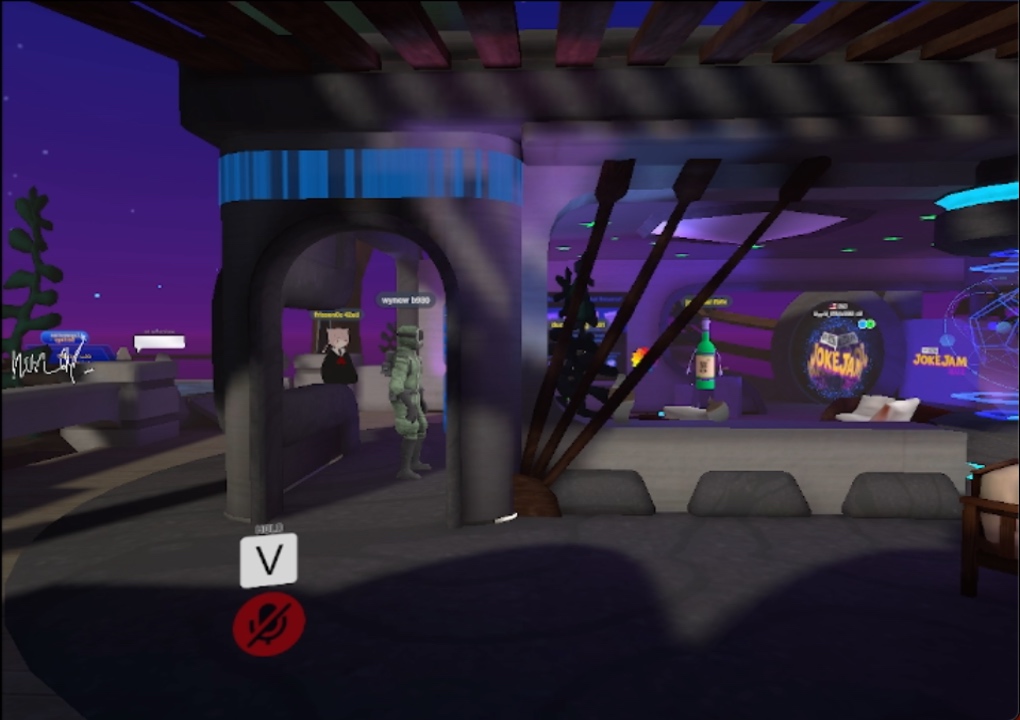}
	\caption{HC3}
\end{subfigure}
	\begin{subfigure}[b]{0.19\textwidth}
	\centering
	\includegraphics[width=0.99\linewidth]{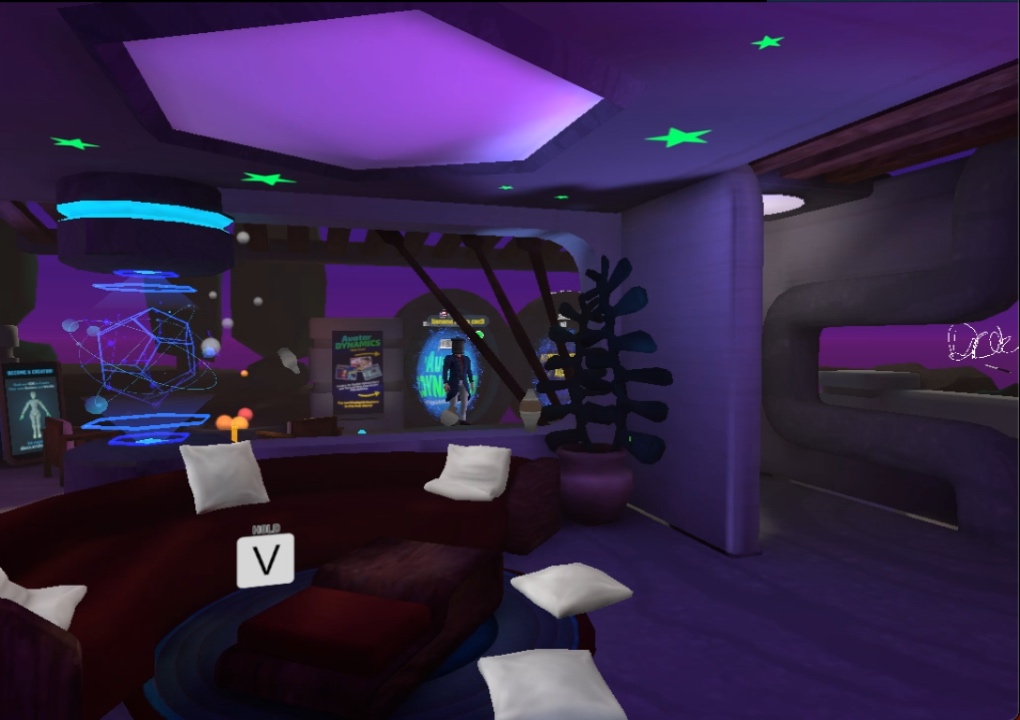}
	\caption{HC4}
\end{subfigure}
	\begin{subfigure}[b]{0.19\textwidth}
	\centering
	\includegraphics[width=0.99\linewidth]{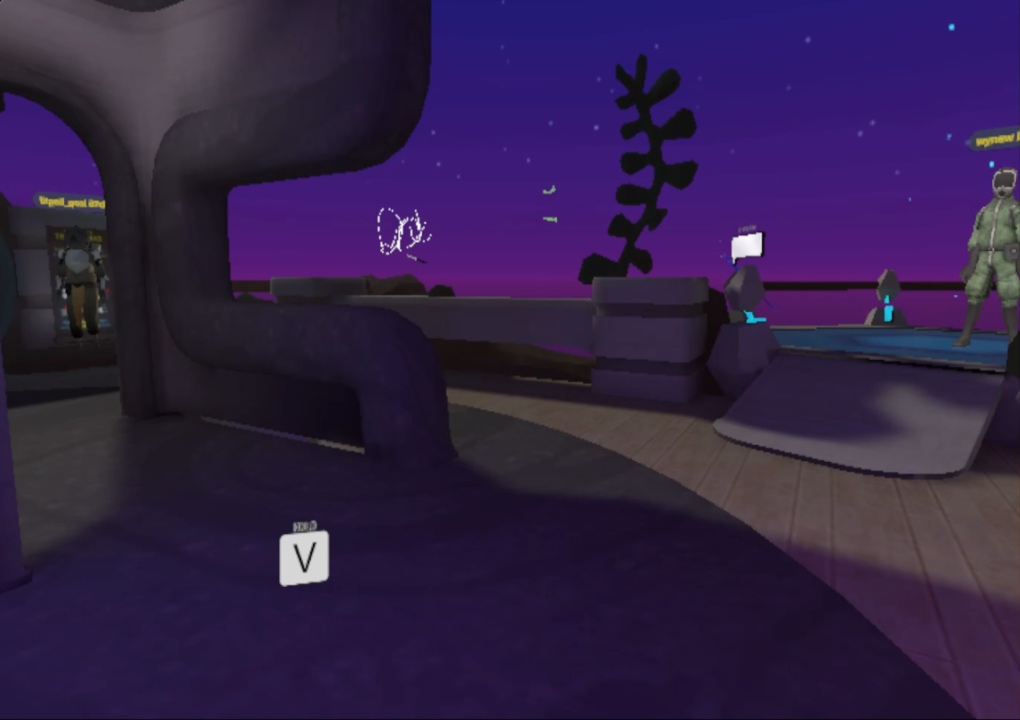}
	\caption{HC5}
\end{subfigure}
	\end{subfigure}

	\begin{subfigure}[b]{0.99\textwidth}
	\centering
	\begin{subfigure}[b]{0.19\textwidth}
		\centering
		\includegraphics[width=0.99\linewidth]{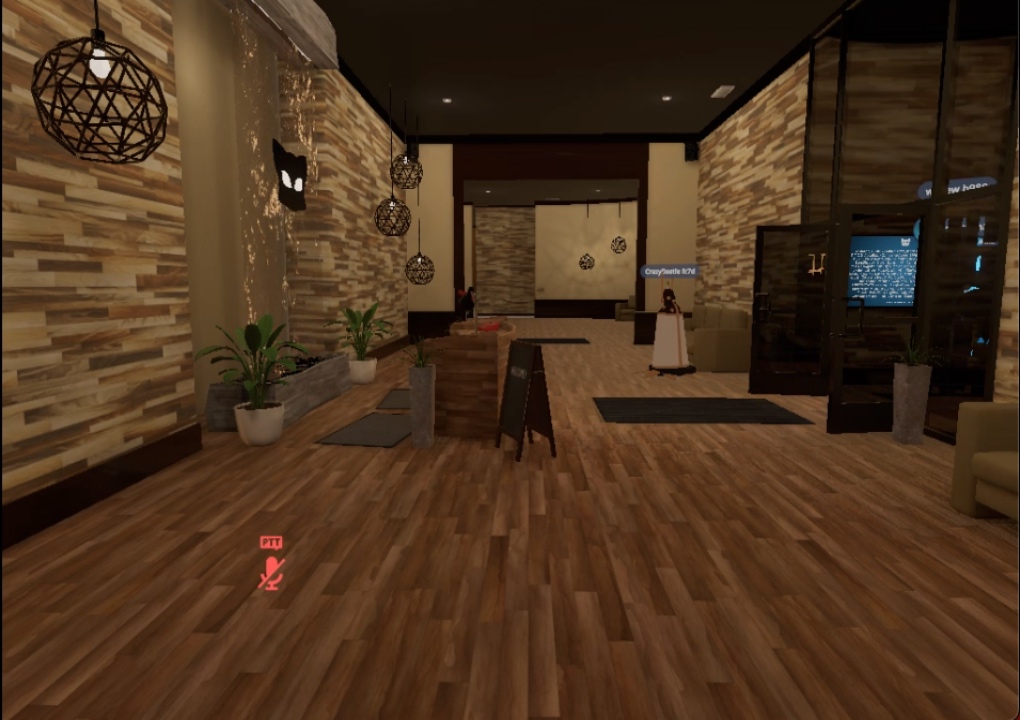}
		\caption{BC1}
	\end{subfigure}
	\begin{subfigure}[b]{0.19\textwidth}
		\centering
		\includegraphics[width=0.99\linewidth]{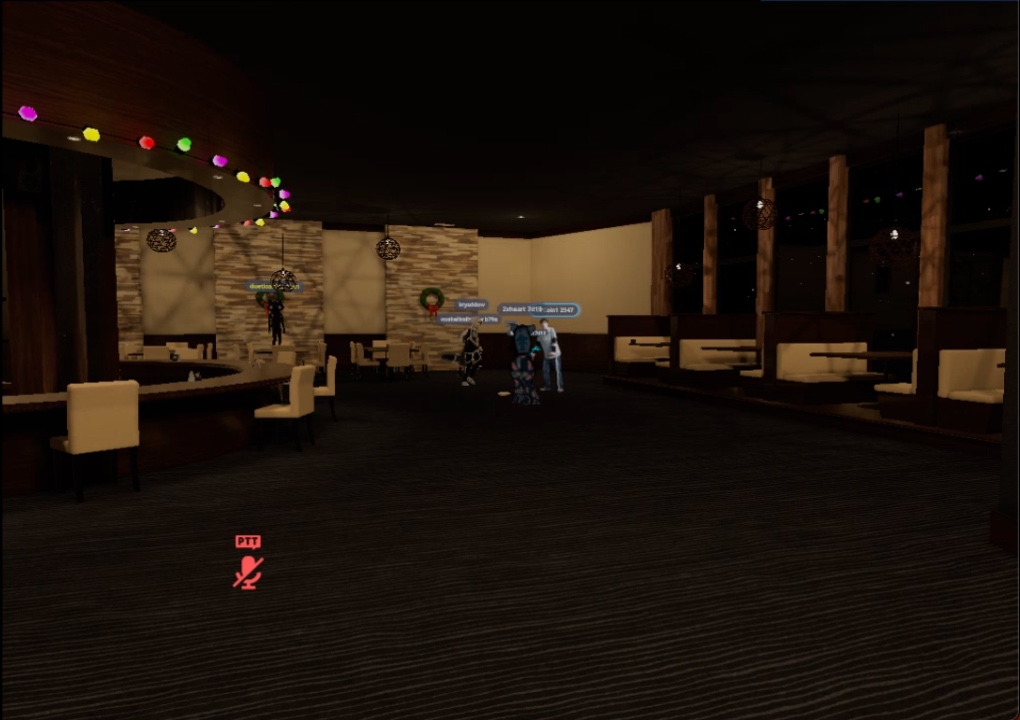}
		\caption{BC2}
	\end{subfigure}	\begin{subfigure}[b]{0.19\textwidth}
		\centering
		\includegraphics[width=0.99\linewidth]{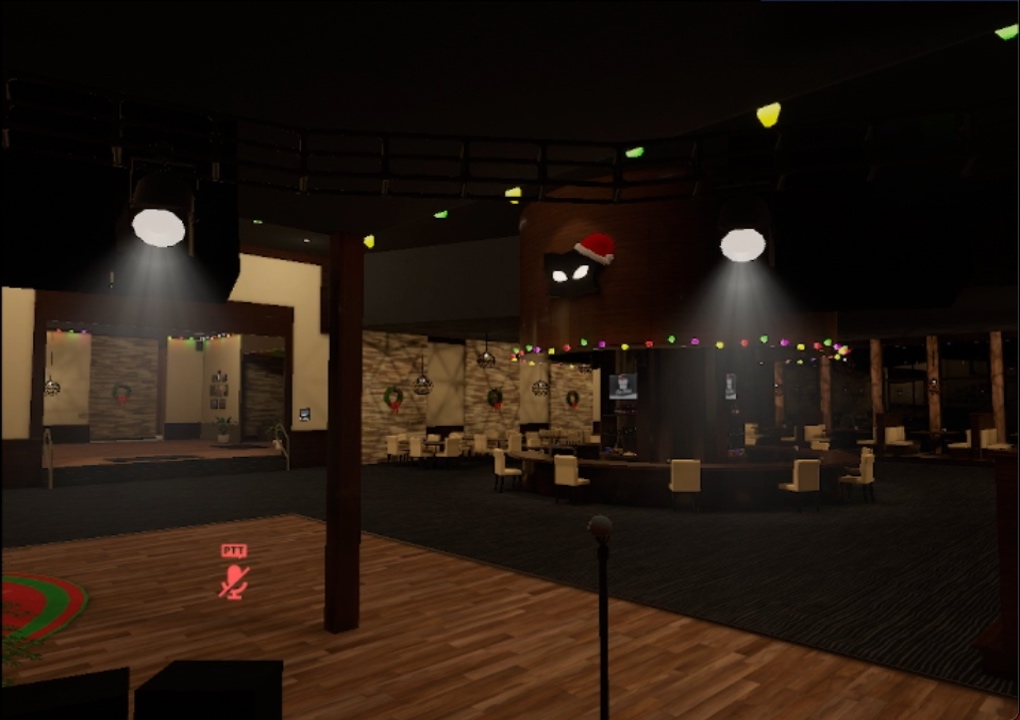}
		\caption{BC3}
	\end{subfigure}
	\begin{subfigure}[b]{0.19\textwidth}
		\centering
		\includegraphics[width=0.99\linewidth]{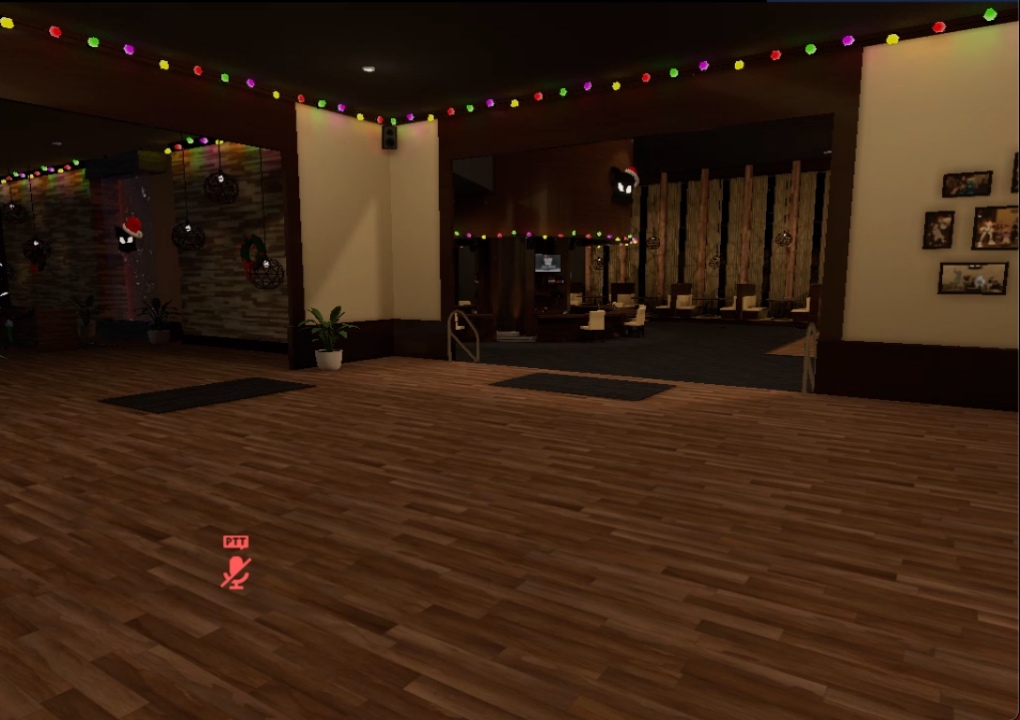}
		\caption{BC4}
	\end{subfigure}
	\begin{subfigure}[b]{0.19\textwidth}
		\centering
		\includegraphics[width=0.99\linewidth]{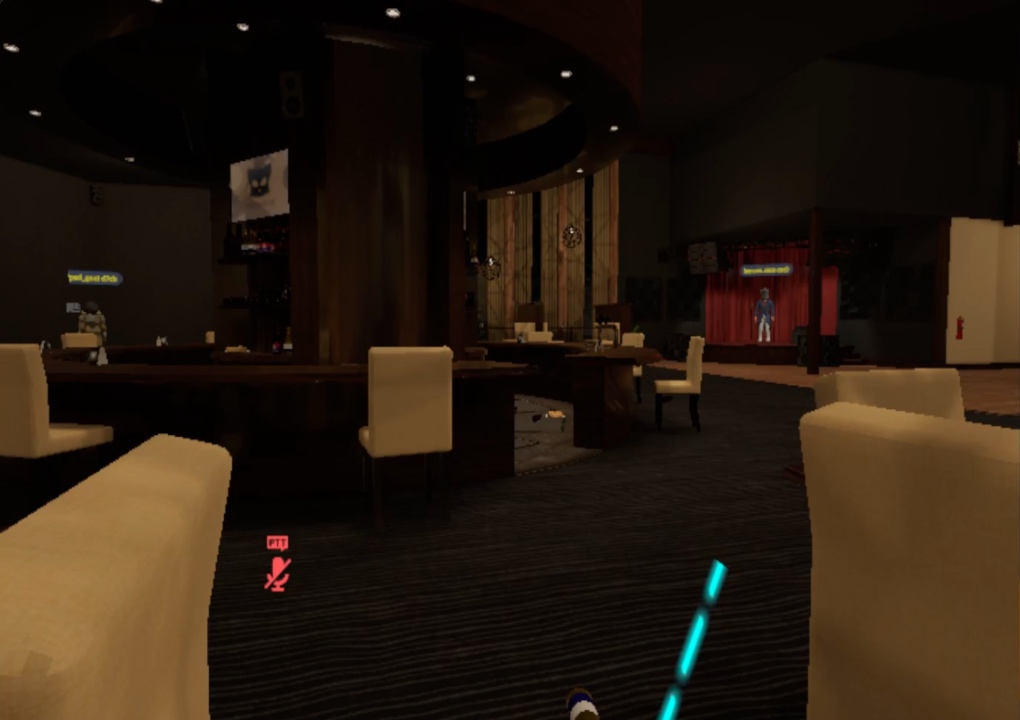}
		\caption{BC5}
	\end{subfigure}	
\end{subfigure}
	\caption{Adversarial viewpoints.}
	\label{fig:adversarialPerspectives}	
\end{figure*}

\section{Experimental Setup}
\label{sec:setup}

To evaluate our proposed correlation framework and training methodology, we collect test (visual and motion sensor) data from human subject participants using a real VR application. In this section, we outline the details of our data collection procedure.

\subsection{Participants' Task}
\label{subsec:workflow}
Our participants (details in \cref{subsec:participants}) carry out a set of representative activities in a virtual reality app while carrying a smartphone and smartphone on their body (details in \cref{subsec:apparatus}). \cref{tab:activities} details all the different types of activities that participants were instructed to perform, in addition to other uncontrolled activities that they may perform while navigating inside the virtual world. The controlled actions include movement of the head, arms, palms, legs, and also actions that require combination of them. These different actions were chosen to generate a variety of different movements within our limited time with the participants. During the uncontrolled activity phases, participants were free to interact with the VR app on their own volition, not limited by the aforementioned activities.
The average time our participants spent on the virtual reality app, in order to provide us data for our study, was 1 hours and 8 minutes. %

\subsection{Adversarial Viewpoints}
\label{subsec:perspectives}
We continuously observe and record the participants' avatar (\cref{fig:adversarialPerspectives}) in the virtual world by means of five different virtual camera positions, where each camera position represents a different adversarial viewpoints. Four of these camera positions are static and positioned at different corners of the virtual room (\cref{fig:adversarialPerspectives}), each of which represents the fixed (or static) position of an adversarial avatar observing the target participant from that position. The fifth camera is mobile, and represents the view of an adversarial avatar moving and navigating in the proximity of the (target) participant's avatar. We carried out our experiments in two different virtual worlds -- one in a public world (called Black Cat) where other real users' avatars may be present, and second in a private world (called Home) where access is restricted to a select group of users. We refer to these five adversarial viewpoints in these two worlds by means of a legend outlined in \cref{tab:cameralegends}. In our evaluation (\cref{sec:eval}), we will also analyze the effect of combining these five viewpoints on the accuracy of activity classification (where the viewpoints are referred to as HCC and BCC for Home and Black Cat, respectively).

\begin{table}[t]
    \centering
    \small
    \caption{Legend of camera viewpoints used in \cref{sec:eval}.}
      \begin{tabular}{cc|cc}
      \toprule
      Home & Legend & Black Cat & Legend \\
      \midrule
      Static Camera 1 & HC1   & Static Camera 1 & BC1 \\
      Static Camera 2 & HC2   & Static Camera 2 & BC2 \\
      Static Camera 3 & HC3   & Static Camera 3 & BC3 \\
      Static Camera 4 & HC4   & Static Camera 4 & BC4 \\
      Mobile Camera & HC5   & Mobile Camera & BC5 \\
      Combined & HCC   & Combined & BCC \\
      \bottomrule
      \end{tabular}%
    \label{tab:cameralegends}%
	\vspace{-0.1in}
\end{table}%

\begin{table}[t]
    \centering
    \caption{Background details of the 35 participants.}
      \begin{tabular}{cc}
    \hline
      \multicolumn{2}{c}{\textbf{Gender}} \\
      14 Female & 21 Male \\
    \hline
      \multicolumn{2}{c}{\textbf{Dominant Hand}} \\
      2 Left & 33 Right \\
    \hline
      \multicolumn{2}{c}{\textbf{VR Familiarity}} \\
      11 Slightly & 24 Moderately-Extremely \\
    \hline
      \multicolumn{2}{c}{\textbf{Prior VR Experience}} \\
      5 Never Used VR Before & 30 Used VR Before \\
    \hline
      \end{tabular}%
    \label{tab:demography}%
	\vspace{-0.1in}
\end{table}%

\subsection{Participants}
\label{subsec:participants}
Between August and December of 2022 we recruited 64 participants for test data collection. However, due to various personal, technical, and medical factors, only 35 of them completed the study and whose data is included in our evaluation. Participants aged between 18 and 48, with a median age of 19. 
Additional demographic  and other details about our participants are listed in \cref{tab:demography}. %
All participants were appropriately compensated for their time and our study procedure was approved by our university's Institutional Review Board (IRB).

\subsection{Data Collection Apparatus}
\label{subsec:apparatus}

\noindent
\textbf{VR Device and App.}
We utilize the Meta Quest 2 VR device\footnote{\url{https://www.meta.com/quest/products/quest-2}} and the popular VRChat \cite{vrchat} app (installed on the Quest 2) for generating and collecting test data from the participants in our study.
As of July 2022, VRChat had more than 200,000 daily active users and more than 7 million registered users \cite{VRChatMMOStats}. 
Moreover, VRChat was one of the few VR apps which supported full-body avatars (instead of only the upper body) at the time we started our experiments. 
Although other popular apps later added integration of full-body avatars \cite{metaverse-legs}, the fundamental nature of data generation (and collection) does not significantly differ across a majority of the VR apps.

\noindent
\textbf{Motion Sensors.}
Participants' body motion was captured at 20 $ms$ sampling interval on a smartwatch (TicWatch 2) worn by the participants on their wrist and on a smartphone (Moto G7 Play) placed in their pocket. 10 participants chose to wear the smartwatch on their right wrist, while the rest chose to wear it on their left wrist. 23 participants placed the smartphone in one of their front pockets, while the rest place it in one of their back pockets.

\noindent
\textbf{Data Logging.}
The VRChat app was installed on five different desktops to record the viewpoints/perspective of an adversary as described in \cref{sec:threatmodel}, and OBS Studio \cite{obsproject} was used to record the each adversarial perspective into individual video files with timestamps. The motion sensors were logged in respective devices with timestamps, and later transferred to another desktop for analysis.

\noindent
\textbf{Analysis Computer.}
A 2021 MacBook Pro was used to train and classify activities, and also for the activity-based filtering and magnitude-based rankings. It is equipped with 10-Core M1 CPU, 16-Core GPU, 16GB memory, 1TB SSD storage, and 16-core Neural Engine. For our \emph{large-scale} analysis in \cref{eval-scalability}, we also used a desktop with Ryzen 5 3600 6-Core 3.6GHz CPU, RTX 3060 12GB GPU, 1TB SSD storage, and 16GB memory, to train and generate large datasets using CTGAN \cite{Diyago2020tabgan,xu2019modeling}.

\begin{figure*}[]
	\begin{subfigure}[b]{0.19\textwidth}
		\centering
		\includegraphics[width=0.99\linewidth]{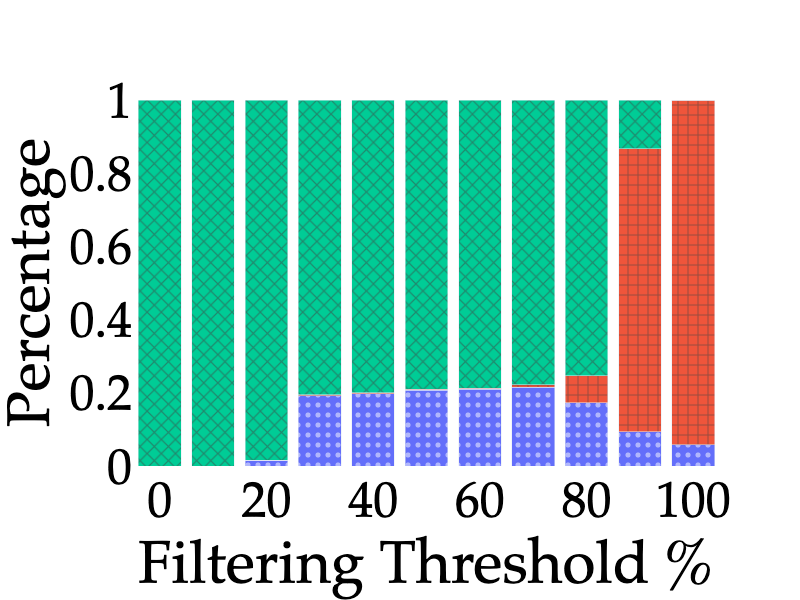}
		\caption{$w = 0.5 s$}
		\label{fig:w0.5}
	\end{subfigure}
	\begin{subfigure}[b]{0.19\textwidth}
		\centering
		\includegraphics[width=0.99\linewidth]{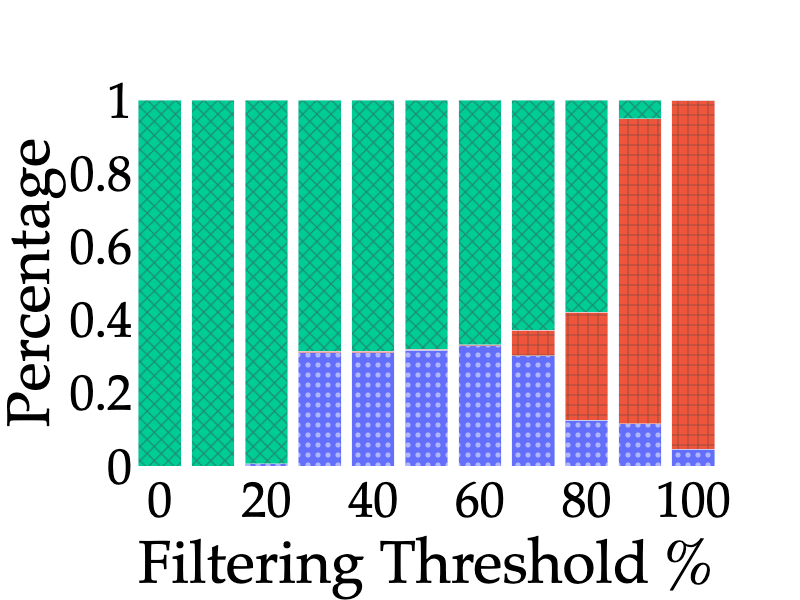}
		\caption{$w = 1 s$}
		\label{fig:w1}
	\end{subfigure}%
\begin{subfigure}[b]{0.19\textwidth}
		\centering
		\includegraphics[width=0.99\linewidth]{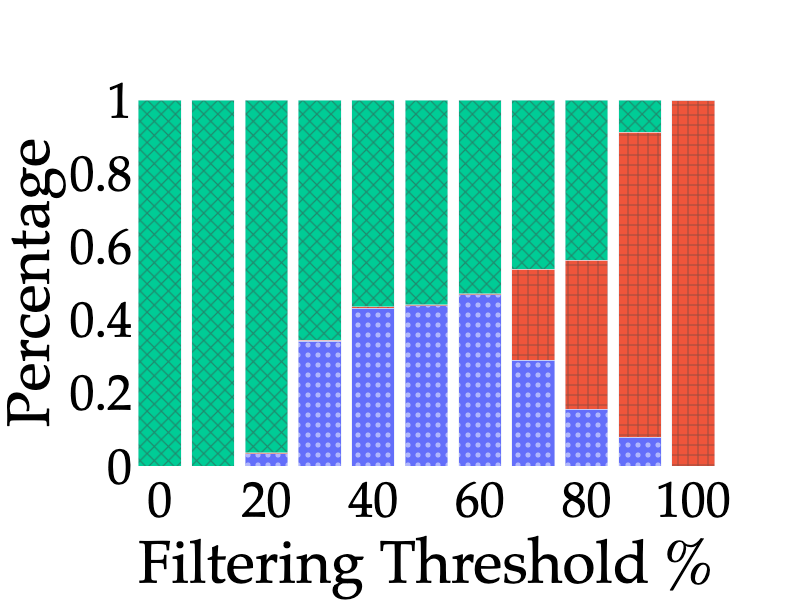}
		\caption{$w = 2 s$}
		\label{fig:w2}
	\end{subfigure}
\begin{subfigure}[b]{0.19\textwidth}
	\centering
	\includegraphics[width=0.99\linewidth]{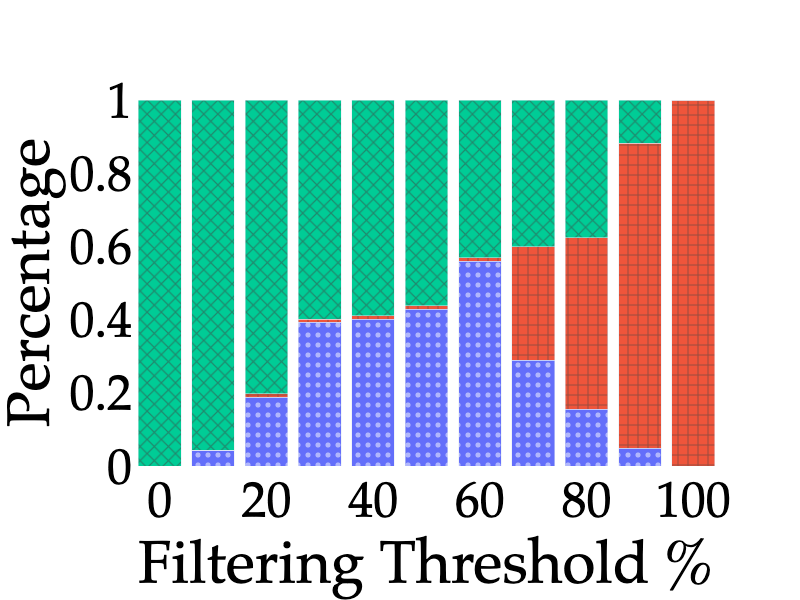}
	\caption{$w = 3$}
	\label{fig:w3}
	\end{subfigure}%
	\begin{subfigure}[b]{0.19\textwidth}
	\centering
	\includegraphics[width=0.99\linewidth]{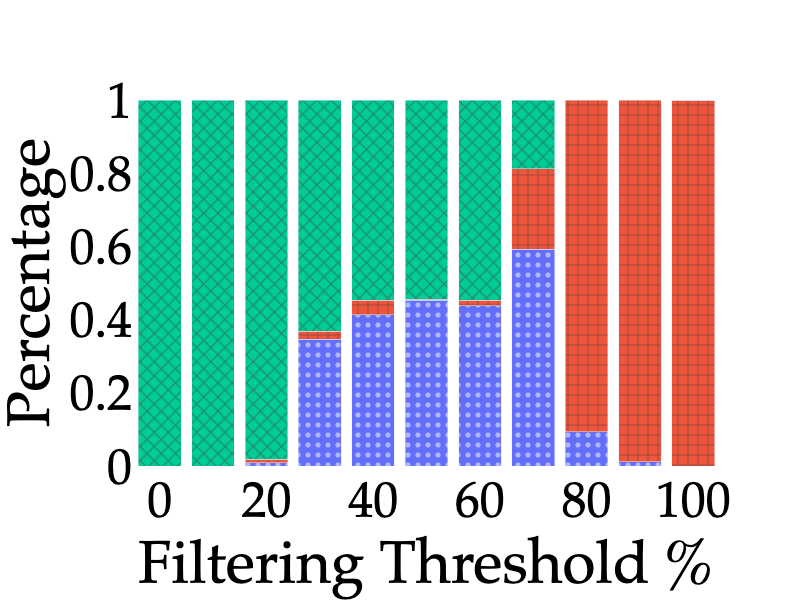}
	\caption{$w = 5 s$}
	\label{fig:w5}
	\end{subfigure}
	
	\vspace{1em}
	\begin{subfigure}[b]{0.99\textwidth}
		\centering
		\includegraphics[width=0.49\linewidth]{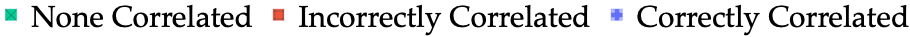}
	\end{subfigure}
	
\caption{Right smartwatch motion sensor and visual movement data correlated with different $w$ and normalized $t$ parameters. Accuracy based on top-1 identity in the rankings. %
} 
\label{fig:rww}
\end{figure*}

\begin{figure*}[]
	\begin{subfigure}[b]{0.19\textwidth}
		\centering
		\includegraphics[width=0.99\linewidth]{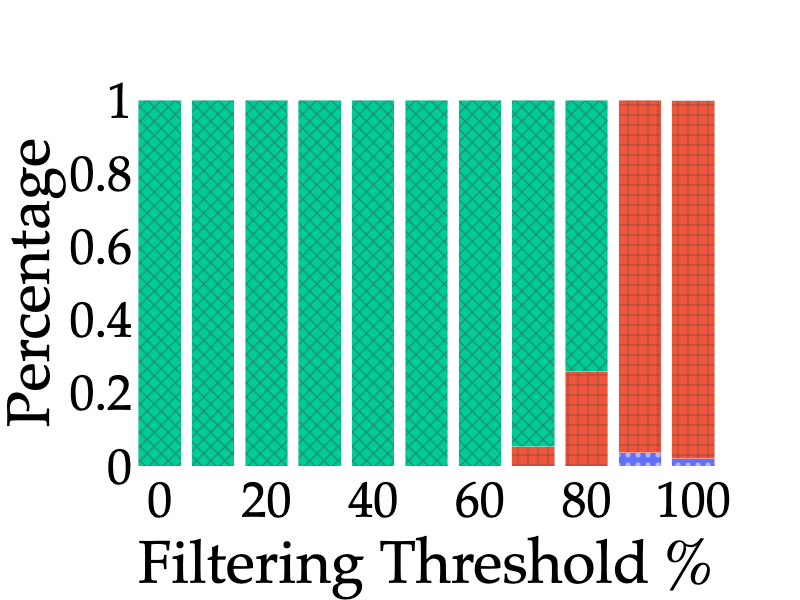}
		\caption{$w = 0.5 s$}
		\label{fig:frw0.5}
	\end{subfigure}
	\begin{subfigure}[b]{0.19\textwidth}
		\centering
		\includegraphics[width=0.99\linewidth]{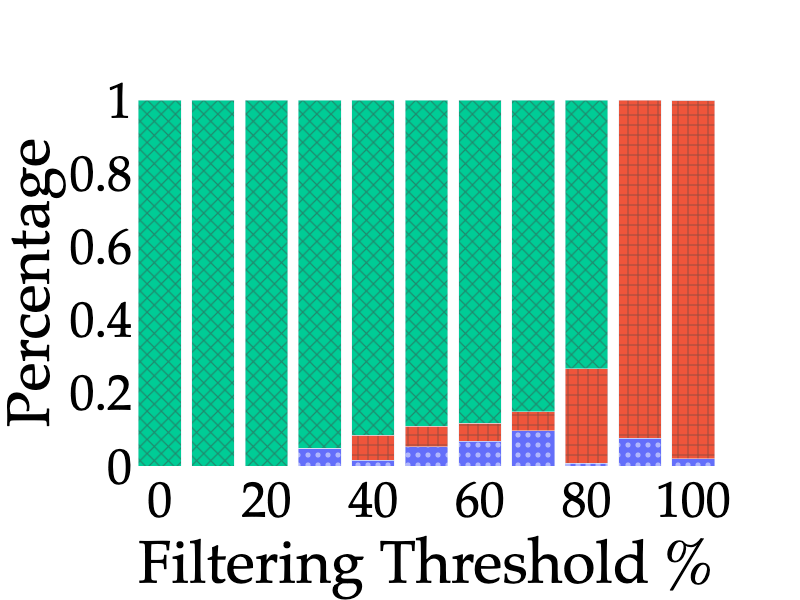}
		\caption{$w = 1 s$}
		\label{fig:frw1}
	\end{subfigure}%
	\begin{subfigure}[b]{0.19\textwidth}
		\centering
		\includegraphics[width=0.99\linewidth]{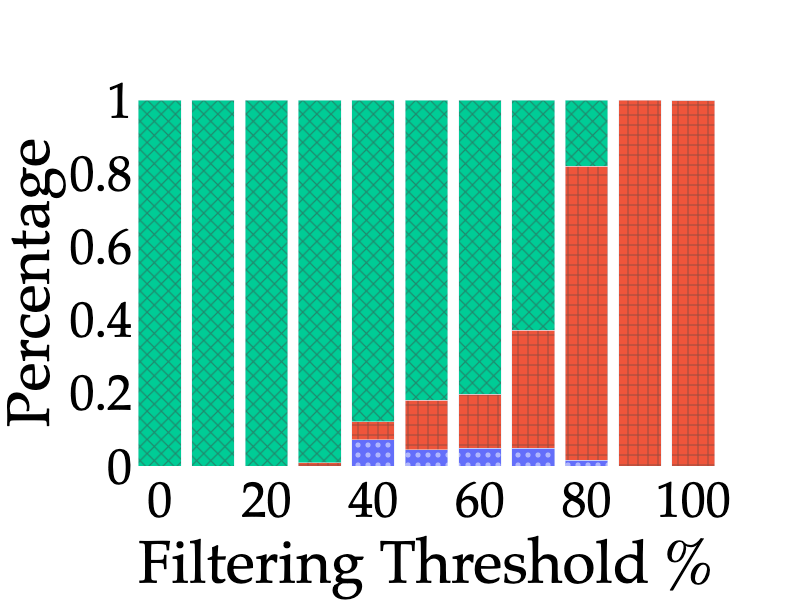}
		\caption{$w = 2 s$}
		\label{fig:frw2}
	\end{subfigure}
	\begin{subfigure}[b]{0.19\textwidth}
		\centering
		\includegraphics[width=0.99\linewidth]{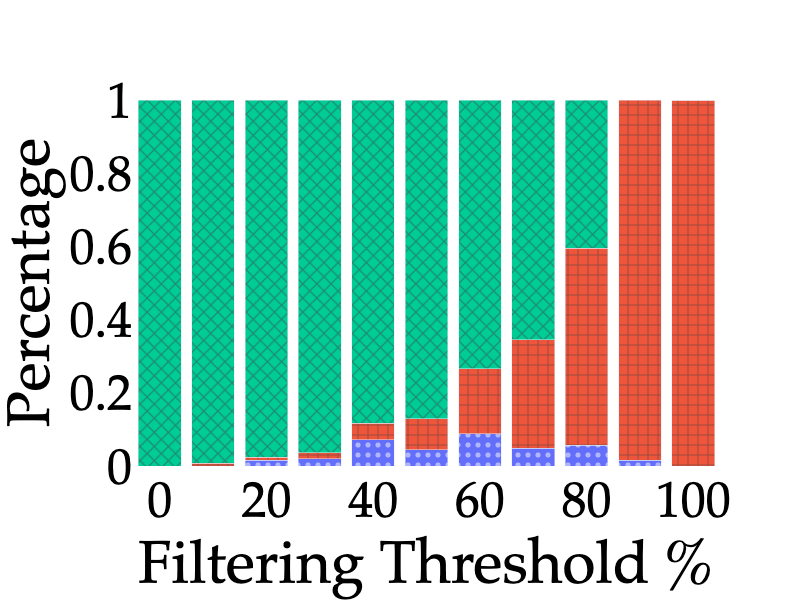}
		\caption{$w = 3 s$}
		\label{fig:frw3}
	\end{subfigure}%
	\begin{subfigure}[b]{0.19\textwidth}
		\centering
		\includegraphics[width=0.99\linewidth]{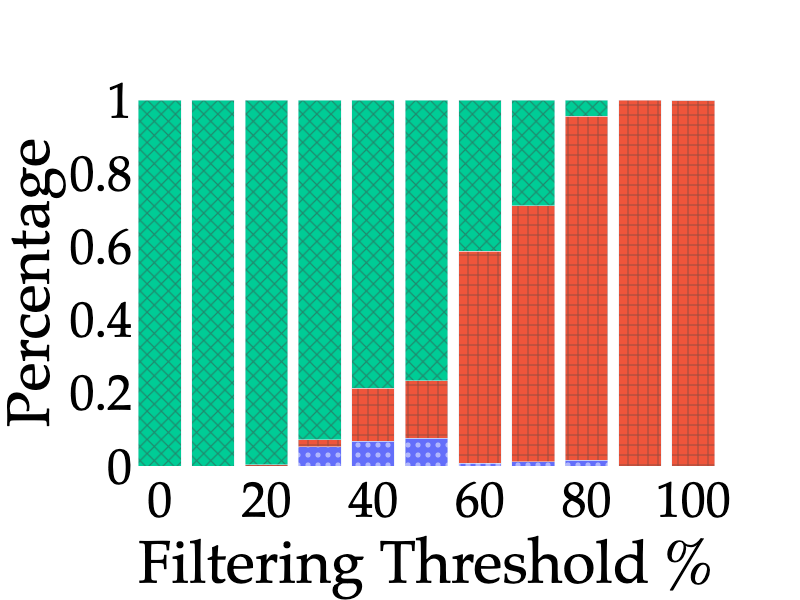}
		\caption{$w = 5 s$}
		\label{fig:frw5}
	\end{subfigure}
	
	\vspace{1em}
	\begin{subfigure}[b]{0.99\textwidth}
		\centering
		\includegraphics[width=0.49\linewidth]{fig-legend}
	\end{subfigure}
	\caption{Front right pocket smartphone motion sensor and visual movement data correlated with different $w$ and normalized $t$ parameters. Accuracy based on top-1 identity in the rankings.}
	\label{fig:frw}
\end{figure*}

\section{Evaluation}
\label{sec:eval}
We next evaluate the proposed correlation framework utilizing the test data collected from participants, which represents a \emph{small-scale} attack with anonymity set size of 271 (accumulating different motion sensor locations from individual participants). We start with identifying suitable framework parameter values such as the activity window size ($w$) and activity-based filtering threshold ($t$).
After extensively evaluating the correlation framework in the \emph{small-scale} setting, we also generate and evaluate a representative dataset for a \emph{large-scale} correlation in \cref{eval-scalability}.

\label{sec:txtStartsHere}
\subsection {Framework Parameters}
\label{eval-parameterThresholds}

Our correlation framework has two key parameters that are critical for the rest of our empirical evaluation. The first parameter is the activity window size ($w$), which is the time duration used to classify an action. The second parameter is the Hamming distance used as the activity-based filtering threshold ($t$), which is the minimum requirement for an activity-vector to be considered in the identity ranking. 
As the total observation time, and thus the number of observed activity windows, will vary between different target users, the activity-based filtering threshold ($t$) is normalized with respect to the number of observed activity windows. No filtering occurs when the filtering threshold is set at 100\%, whereas at 0\% even one mismatch in the activity sequence will result in that activity-vector being filtered out.

\cref{fig:rww,fig:frw} show the correlation accuracy, where "None Correlated" occurs when the activity-based filtering filters all candidate activity-vectors, "Incorrectly Correlated" occurs when the top ranked identity is incorrect, and "Correctly Correlated" occurs when the top ranked identity is correct.
From these figures we can see an overall trend that as we increase $w$, the percentage of identities that passes the activity-based filtering and then used for identity ranking also grows. Conversely, the percentage of ``None Correlated'' is diminished as $w$ is increased. This can primarily be attributed to (i) the size of activity sequence in the activity-vector is inversely proportional to $w$ for a constant observation time period thereby reducing the number of probable mismatches, and (ii) the activity inference tends to perform more accurately for larger $w$.

While the above observation should compel us to select a larger $w$, in \cref{fig:rww,fig:frw} we also observe that there exists a trade-off between $w$ and correctly correlated identities for different activity-based filtering thresholds. For instance, when $w=5 s$ we observe that the percentage of correctly correlated identities starts to decrease beyond the filtering threshold of 70\% in \cref{fig:w5}. This is most likely because as the size of activity-vector is reduced with increasing $w$, the probability of confusion with another person's activity magnitudes is increased. This trend was consistent across other experimental variables, such as different adversarial viewpoints, different motion sensors, and different motion sensor positions on the body. 

\emph{Based on empirical observations across different experimental variables we set $w=1 s$ and $t = 30\%$ for the rest of our analyses}. On average, these selected values are best suited for maximizing the percentage of correctly correlated identities. The average correctly correlated identities using these parameter values within top-1 of the ranking was 16.3\%, and 17.0\% of the identities were within top-3. %
In an alternate adversarial model where the motion sensor positions on the body is known to the adversary, more specific (i.e., per target user) $w$ and $t$ values can be selected to further improve the percentage of correctly correlated identities.

\subsection {Activity Confusions}
\label{eval-actionClassificationConfusion}

\begin{figure}[t]
	\centering	
		\begin{subfigure}[b]{0.49\textwidth}
		\centering
		\includegraphics[width=0.99\linewidth]{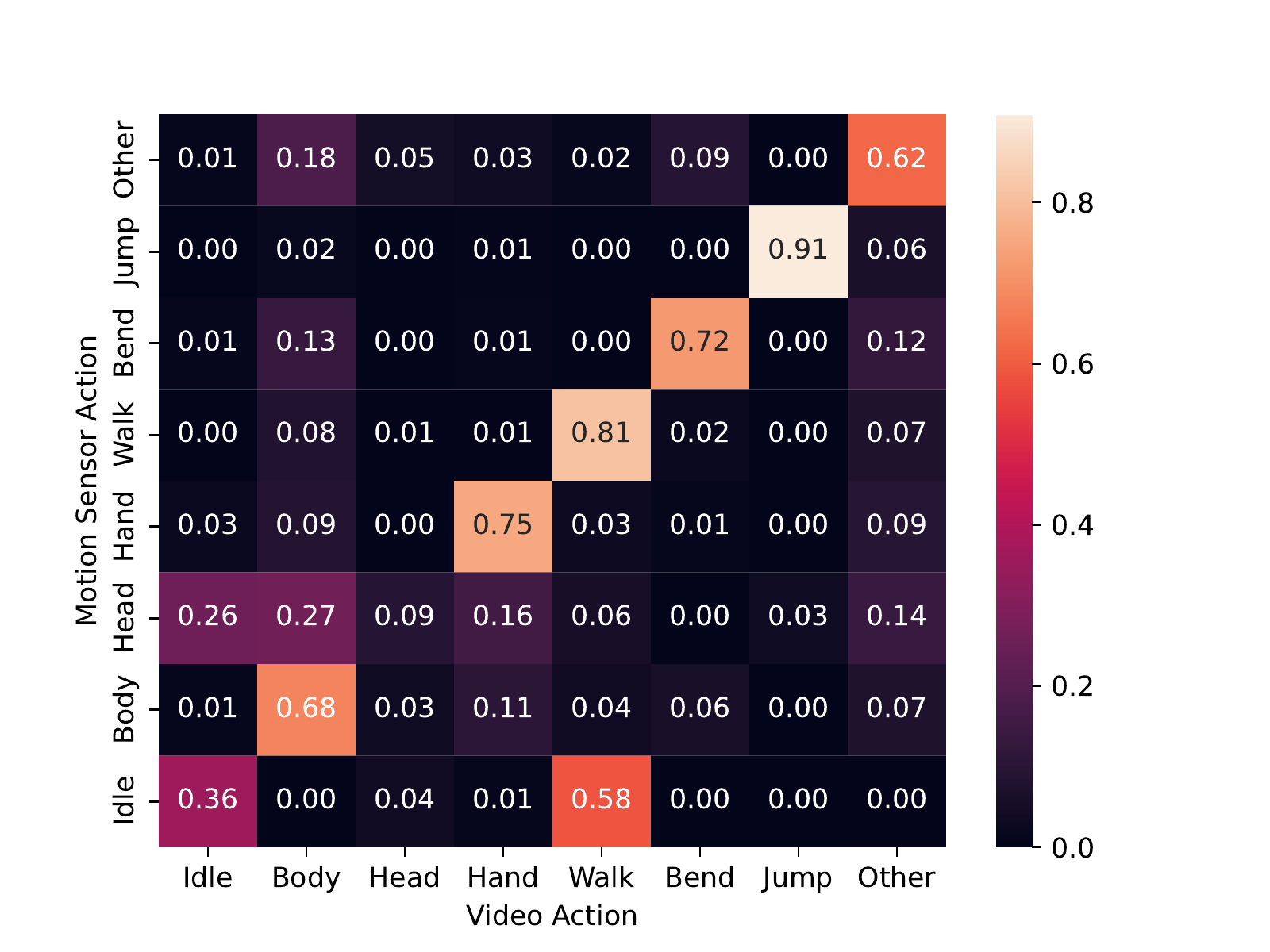}
		\caption{Using right wrist smartwatch.}
		\label{fig:aca}
	\end{subfigure}
	\begin{subfigure}[b]{0.49\textwidth}
		\centering
		\includegraphics[width=0.99\linewidth]{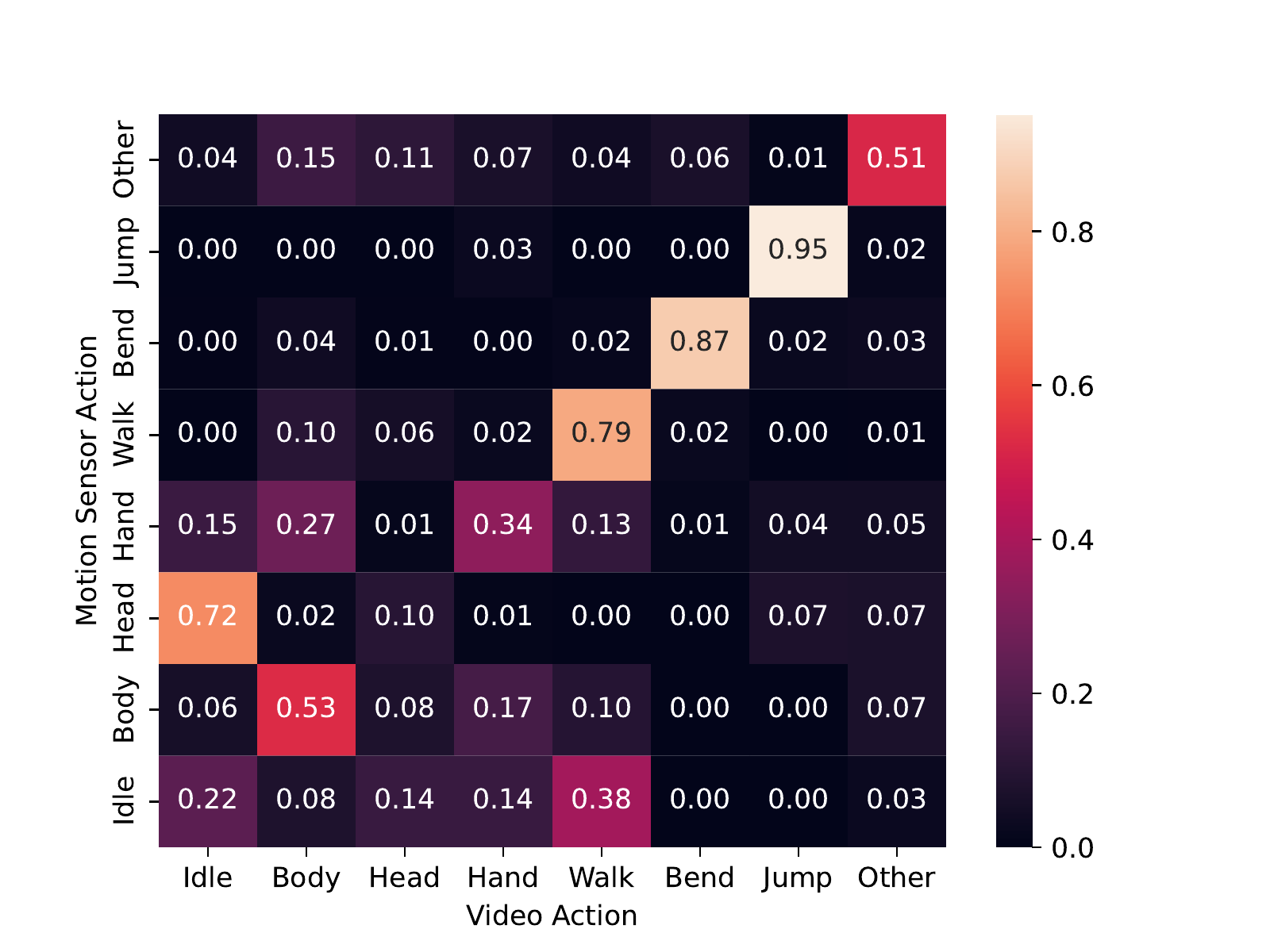}
		\caption{Using front right pocket smartphone.}
		\label{fig:acb}
	\end{subfigure}%
\caption{Activity classification confusion between motion sensor data and visual movements.}
\label{fig:actionsConfusion}
\vspace{-0.15in}
\end{figure}

\begin{figure}[t]
	\centering
	\includegraphics[width=0.55\linewidth]{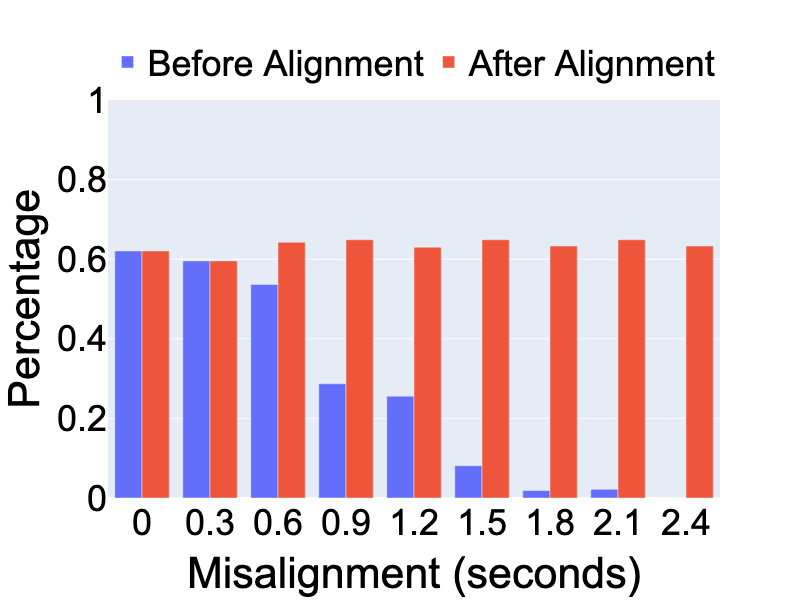}
	\caption{Correctly correlated accuracy (top-1 rank) with artificially introduced misalignment, shown for data from the right wrist.%
	}
	\label{fig:missSync}
	\vspace{-0.15in}
\end{figure}

The accuracy of the activity classification models play an important role in the correlation framework's overall success rate. Activity classification between visual and motion sensor data differs significantly due to the modality (of input signal), and is potentially subject to different types of noises and interference signals.
Different adversarial viewpoint angles, distances, and occlusion levels affect the visual data classification. For instance, if only half of the avatar is visible due to being behind a coach or another avatar is in front of the target avatar, the chance of a misclassification is significantly increased.
On the other hand, the positioning and orientation of the device used to collect motion sensor data also imposes certain limitations on the activity classification accuracy, especially as we assume that the adversary is unaware of the exact position of the motion sensor. For instance, if the motion sensor data is from a smartwatch worn on the right hand, it is very useful to classify activities involving the right hand, but may result in high misclassification of activities not involving the right arm. 

Due to these apparent limitations, we analyze the direct consequence of misclassifications, i. e., the confusion of activities between the visual and motion sensor data. In \cref{fig:actionsConfusion}, we observe that the \emph{idle} activity has noticeably low accuracy (36\% and 22\% for right wrist smartwatch and front right pocket smartphone, respectively), and is often confused with other activities.
An unexpected, yet clearly discernible, confusion exists between motion sensor \emph{idle} and visual \emph{walking}.
One possible factor behind this observation is that VR users may be using the VR joystick to walk in the virtual world. As a result, the target user appears idle in the motion sensor data, while their virtual avatar is visually walking. 
Another noteworthy observation is that head movements had high confusion due to the fact that placement of motion sensors around hip and wrist areas is not suitable for capturing the target user's head movements, where as a head-mounted VR device is accurately able to capture head movements and apply them to the avatar in the virtual world. %

In light of these insights, we further optimize our framework as follows. Rather than considering all the classified actions, we only utilize activities with less than 60\% of confusion -- body, hand, walk, bend, jump, and others -- for our activity-based filtering. %
Remaining activities in the activity-vector are ignored from the Hamming distance calculations. The average correctly correlated identities after this optimization within top-1 of the ranking was 37.3\%, while 38.7\% of the identities were within top-3. %

\subsection {Time Alignment}
\label{eval-unsyncData}

Both the visual and motion sensor data are collected with device timestamps for synchronization. Although most modern smartphones and smartwatches are by default periodically updated against internet-based time servers, motion sensor data collection in the wild may contain time drift errors and thus misaligned with the visual movements. %
Misaligned data sources will likely cause confusion between classified activities, resulting in a high failure rate in satisfying the activity-based filter threshold. As shown in \cref{fig:missSync}, misaligned data can drop a $62.1\%$ correctly correlated result down to $0\%$ in the presence of only 2.4 seconds (of artificially introduced) misalignment. The adversary can potentially detect and overcome such misalignments by offsetting the (motion sensor) data in increments, and selecting a time offset ($\pm\delta$) that results in the minimum Hamming distance in the activity-based filtering. Realistic assumptions must be made on the bounds of $\delta$ in order to keep the computational time practical.

\begin{figure*}[]
	\begin{subfigure}[b]{0.33\textwidth}
		\centering
		\includegraphics[width=0.99\linewidth]{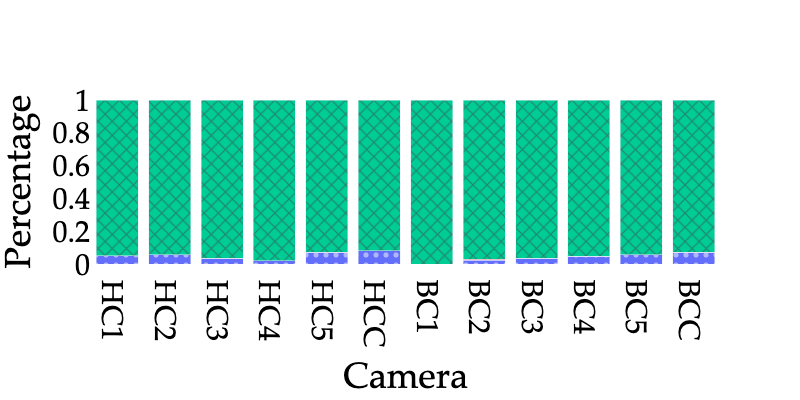}
		\caption{Motion sensor in back left pocket}
		\label{fig:backLeftSensor}
	\end{subfigure}
\begin{subfigure}[b]{0.33\textwidth}
		\centering
		\includegraphics[width=0.99\linewidth]{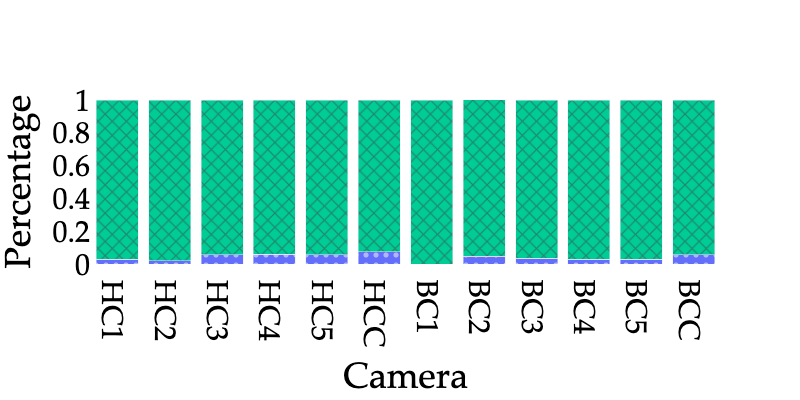}
		\caption{Motion sensor in back right pocket.}
		\label{fig:backRightSensor}
	\end{subfigure}%
\begin{subfigure}[b]{0.33\textwidth}
		\centering
		\includegraphics[width=0.99\linewidth]{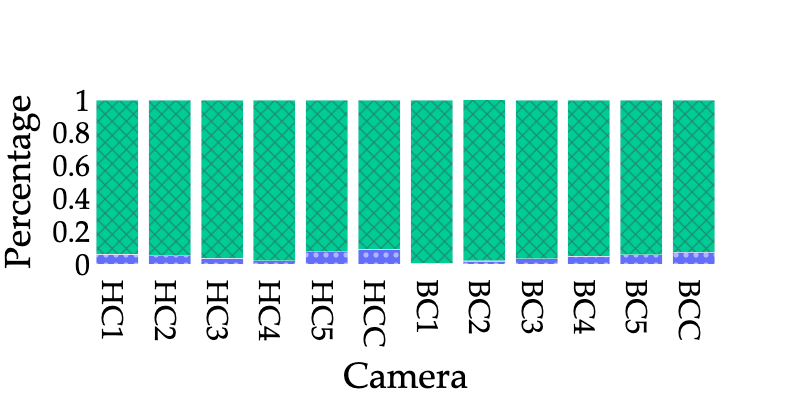}
		\caption{Motion sensor in front left pocket.}
		\label{fig:frontLeftSensor}
	\end{subfigure}
	\begin{subfigure}[b]{0.33\textwidth}
		\centering
		\includegraphics[width=0.99\linewidth]{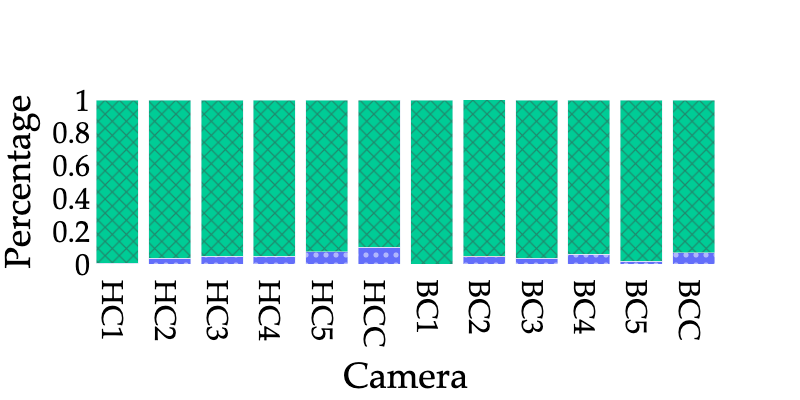}
		\caption{Motion sensor in front right pocket.}
		\label{fig:frontRightSensor}
	\end{subfigure}
\begin{subfigure}[b]{0.33\textwidth}
		\centering
		\includegraphics[width=0.99\linewidth]{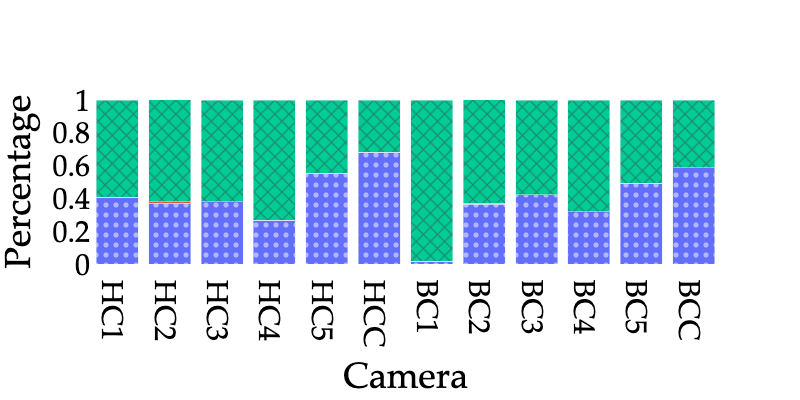}
		\caption{Motion sensor on right wrist.}
		\label{fig:rightWristSensor}
		\label{fig2:tired}
	\end{subfigure}
\begin{subfigure}[b]{0.33\textwidth}
		\centering
		\includegraphics[width=0.99\linewidth]{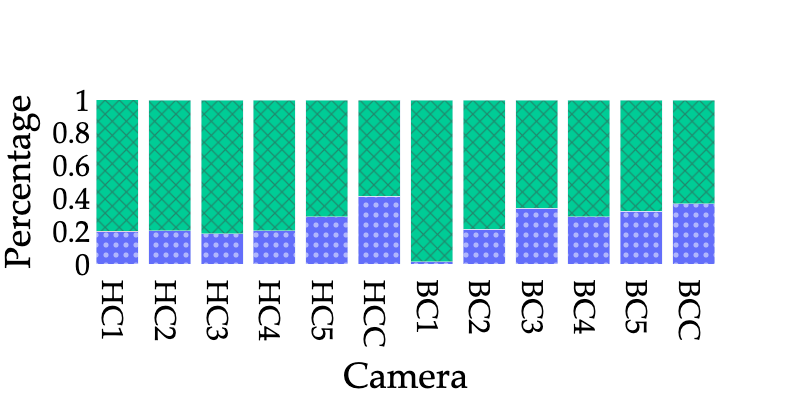}
		\caption{Motion sensor on left wrist.}
		\label{fig:leftWristSensor}
	\end{subfigure}

	\vspace{1em}
\begin{subfigure}[b]{0.99\textwidth}
	\centering
	\includegraphics[width=0.49\linewidth]{fig-legend}
	\label{fig:backLeftSensor}
\end{subfigure}
\caption{Accuracy for different cameras positions and motion sensors locations of devices during the free-movement phase. Accuracy based on top-1 identity in the rankings.}
 \label{fig:Cameras}
\end{figure*}

\subsection {Different Motion Sensors and Camera Locations}
\label{eval-phoneVsWatch}

We next detail how different positions of the motion sensor on the (human) body and different adversarial viewpoints affect the correct correlation of our proposed framework.
Overall, smartwatch (motion sensor) on left or right wrist performed better than the smartphone in the hip pockets (\cref{fig:Cameras}). 
For example, for the \emph{Home} world, the smartwatch yielded about 41\% and 68\% correct correlations (top-1 rank), for left and right wrists, respectively. In contrast, the front left-front pocket smartphone data resulted in about 9.1\% correct correlations, while other smartphone locations are in a similar range. 
Intuitively, one of the main factors behind this observation is the inability of smartphone motion sensors to pick up hand and head movements when they are located in the hip area pockets. This causes higher confusion between activities (\cref{fig:actionsConfusion}), resulting in the activity-vector of the target user being filtered out with high likelihood. 

As far as the impact of different adversarial viewpoints on the correlation accuracy of our framework is concerned, we can see from \cref{fig:Cameras} that, except for BC1, all other camera locations (or adversarial viewpoints) yielded comparable results within each of the motion sensor locations. %
The reason behind BC1 performing particularly poor is that its location was near the entrance point of the \emph{Black Cat} world and most participants eventually moved away from the field-of-view of this camera during the data collection experiments.
In summary, combining multiple viewpoints and the availability of wrist-based motion sensor data are the most favorable conditions for the adversary.

\subsection {Conflicting Activity Sequences}
\label{eval-sameActions}

\begin{figure}[t]
	\centering
		\begin{subfigure}[b]{0.49\linewidth}
			\centering
			\includegraphics[width=0.9\linewidth]{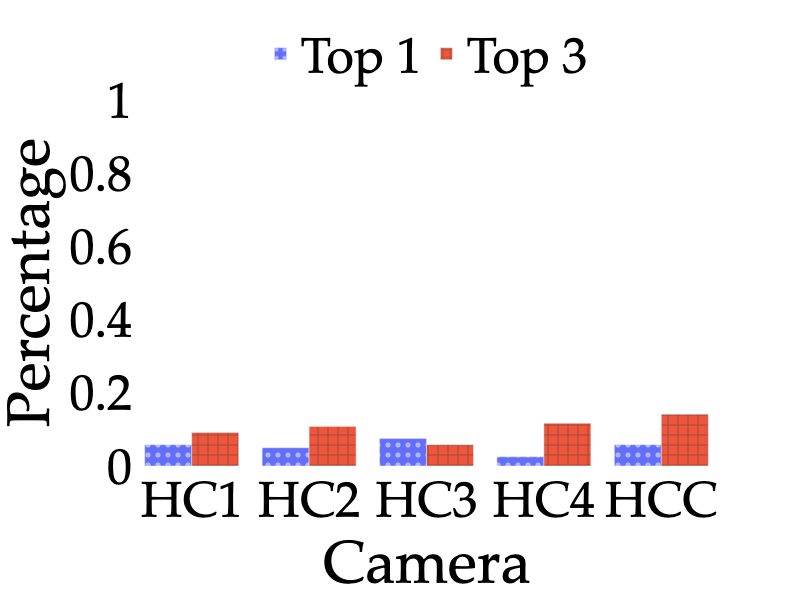}
			\caption{Front left pocket motion data}
			\label{fig:sameFL}
		\end{subfigure}
		\begin{subfigure}[b]{0.49\linewidth}
			\centering
			\includegraphics[width=0.9\linewidth]{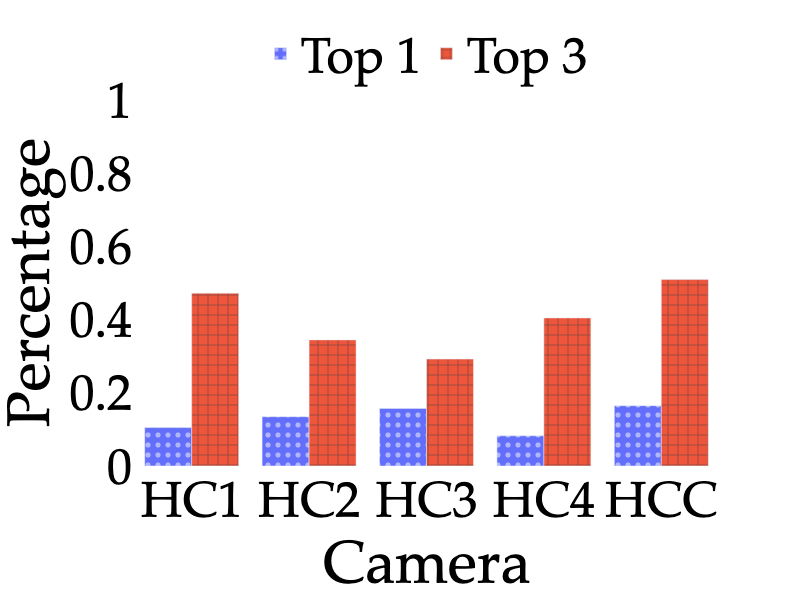}
			\caption{Right 
				wrist motion data}
			\label{fig:sameRW}
		\end{subfigure}
		\caption{Identity correlation for conflicting activity sequences.}
		\label{fig:sameMovement}
		\vspace{-0.15in}
\end{figure}

There can be cases, especially in a \emph{large-scale} attack, where multiple target users perform a similar or even an identical sequence of activities. In such cases, the magnitude-based ranking should ideally still rank the real identity higher than others.
In this part of our analysis, we study the extent to which our magnitude-based ranking is able to do so, by comparing correlation accuracy when participants (and their avatars) performed the same sequence of activities. 
In \cref{fig:sameRW}, we observe $16.5\%$ correct correlation for motion data from the right wrist in top-1 of identity rankings and $50.1\%$ correct correlation within the top-3 ranks. %
This demonstrates that to an extent the magnitude-based ranking is in fact able to discern the difference between identities based on the magnitude of movements.

\begin{table*}[htbp]
  \small
  \centering
  \caption{Computational time and correlation accuracy for GAN generated datasets, for default and optimized activity-based filtering. Tested for $k = 5$ with $t = 2$, and $k = 10$ with $t = 3$. Entries marked as ``--'' did not finish.}
    \begin{tabular}{ccccccc}
    \toprule
    \textit{\textbf{GAN Generated}} & \multicolumn{3}{c}{Time (ms)} & \multicolumn{3}{c}{Average Correctly Correlated (\%)} \\
    \midrule
    Motion $\times$ Video & Default & Hash 3/5 & Hash 7/10 & Default & Hash 3/5 & Hash 7/10 \\
    \midrule
    100 $\times$ 100 & 22    & 7     & 29    & 47.00 & 41.00 & 44.00 \\
    500 $\times$ 500 & 586   & 36    & 177   & 44.00 & 43.00 & 41.00 \\
    1000 $\times$ 1000 & 2415  & 80    & 386   & 45.80 & 39.40 & 40.10 \\
    10000 $\times$ 10000 & 267293 & 1020  & 4714  & 51.30 & 42.60 & 34.70 \\
    100000 $\times$ 100000 & 31481325 & 9617  & 54736 & 39.10 & 31.10 & 21.50 \\
    1000000 $\times$ 1000000 & -     & 99304 & 644596 & -     & 30.90 & 21.90 \\
    \bottomrule
      \end{tabular}%
    \label{tab:scale1}%
  \end{table*}%

\begin{table}[htbp]
  \small
  \centering
  \caption{Correlation accuracy for permutation generated datasets, for default and optimized activity-based filtering. Tested for $k = 5$ with $t = 2$, and $k = 10$ with $t = 3$. Entries marked as ``--'' did not finish.}
    \begin{tabular}{cccc}
    \toprule
    \textit{\textbf{Permutation Generated}} & \multicolumn{3}{c}{Average Correctly Correlated (\%)}  \\
    \midrule
    Motion $\times$ Video & Default & Hash 3/5 & Hash 7/10 \\
    \midrule
    100 $\times$ 100 & 54.00 & 49.00 & 49.00 \\
    500 $\times$ 500 & 55.00 & 42.00 & 40.00 \\
    1000 $\times$ 1000 & 48.90 & 41.10 & 43.10 \\
    10000 $\times$ 10000 & 53.20 & 40.90 & 36.20 \\
    100000 $\times$ 100000 & 51.70 & 36.60 & 29.80 \\
    1000000 $\times$ 1000000 & -     & 38.70 & 24.50 \\
    \bottomrule
      \end{tabular}%
    \label{tab:scale2}%
    \vspace{-0.1in}
  \end{table}%

\section{Optimizing for Large-scale Attacks}
\label{eval-scalability}

An adversary trying to correlate thousands or millions of anonymous avatars with identified motion sensors data is presented with a very significant computational task. In this section, we analyze the computational complexity of this task and propose related optimizations to our correlation framework.

\noindent
\textbf{Synthetic Data Generation}
To test the scalability of our framework, we must first generate a very large synthetic dataset \emph{utilizing} real participant data collected in \cref{sec:setup}.
While it was not feasible for us to collect real-world data from thousands or millions of participants, due to the time and resources required for systematic data collection per participant, we still want to test using a dataset that has resemblance to the \emph{small-scale} dataset instead of generating completely random activity-vectors.
The activity classification and magnitude calculation tasks take constant time, and will grow linearly with the size of each dataset ($p$ and $q$, for visual movement and motion sensor datasets, respectively). For large $p$ and $q$, the more complex task is that of calculating the correlation of all $q$ identities against all $p$ anonymous avatars. %
However, as seen in \cref{sec:eval}, the activity-based filtering is very effective in reducing the complexity of the magnitude-based identity rankings.
Therefore, for large $p$ and $q$ the most computationally complex task in the entire framework comes down to the activity-based filtering. %
Accordingly, we generate our \emph{large-scale} dataset to test the scalability of our activity-based filtering, which only requires activity sequences as input.
Our first \emph{large-scale} dataset was generated using a modern tabular Generative Adversarial Network (GAN) technique \cite{Diyago2020tabgan}, called CTGAN \cite{xu2019modeling}, which is trained using activity sequences from real participants, as outlined in \cref{sec:setup}. Our second \emph{large-scale} dataset was generated using random permutations of our activity sequences from \cref{sec:setup}. 
Each of these \emph{large-scale} datasets contained 1 million activity sequences for the motion sensor and 1 million activity sequences for the visual movements.

\noindent
\textbf{Activity-based Filtering Without Optimizations.}
Without any optimizations, the activity-based filtering has a time complexity of $O(pqk^2)$, where $p$ is the number of unique avatars from the visual movement data, $q$ is the number of different identities from the motion sensor data, and $k$ is the size of the activity sequences. As such, we can further assume that increasing the size of $k$ would have diminishing returns (computationally), making it less attractive for an adversary to record each target for too long. 
Therefore, we assume $k$ would not be scaled, unlike $p$ and $q$, and treat $k$ as constant, thus resulting with a complexity of $O(pq)$. 
As shown in \cref{tab:scale1}, our setup takes $2.2 * 10^1$ ms to finish activity-based filtering when $p = q = 100$. However, when we scale up to $p = q = 10^5$, it requires $3.15 * 10^7$ ms (or about 8 hours) to finish activity-based filtering.
We estimate that for $p = q = 10^6$, it will take approximately 30 days to finish, and about 3000 days when $p = q = 10^7$, which is not very scalable.

\noindent
\textbf{Optimization.}
We propose the use of a hash table to store our activity sequence data in order to reduce the time complexity of activity matching and filtering.
However, as even a single mismatch between two activity sequences will result in completely different hash values (i.e., the keys in a hash table), we design a larger hash table that allows for some degree of mismatch. 
Specifically, we populate a hash table with keys based on permutations of the $q$ activity sequences in $M$ (each of length $k$) from the motion sensors data, accounting for possible errors allowable within the Hamming distance threshold ($t$).  
Let us assume that the numbers $0$ to $7$ denotes 1 of the eight activities we classify. 
If $k = 5$, an example of the activity string would be $\langle47634\rangle$. 
If our hamming distance threshold is $t=2$, then any two activities can be mismatched and still pass the threshold. 
Now, assume the character $*$ as a wildcard activity that may or may not be a match.
To populate the hash table exhaustively, we compute every possible permutation of each activity sequences in $M$ including up to two $*$.
For our previous example, $\langle47634\rangle$, some of the permutations generated would be $\langle**634\rangle$, $\langle4*6*4\rangle$, and $\langle47*3*\rangle$.
All these permutations are then used as the key in our hash table, while the corresponding value is the identity of users from the motion sensor data ($M$).
Thereafter, during the correlation process, each activity sequence from the video dataset also undergoes permutations with up to two $*$, and then queried against the above hash table for a match. If a matching key exists, the corresponding identity and activity-vector has satisfied the activity-based filtering and is included in the identity ranking.

\noindent
\textbf{Optimized Performance Analysis.}
The number of permutations per activity-vector does not scale with the size of datasets and thus can be treated as $O(1)$ time complexity. Similarly, hash table search and insertion is $O(1)$ time complexity. 
Therefore, with the use of our hash table, the new time complexity becomes $O(p+q)$, where $O(q)$ time is required to create the hash table, and $O(p)$ time is require to iterate through $V$ for filtering.
Our empirical results (\cref{tab:scale1,tab:scale2}) show that with the optimization, the activity-filtering is significantly faster. %
For instance with $p = q = 100000$, $k = 10$, and $t = 3$, using the optimization technique was $575$ times faster than the default activity-based filter. %

\section{Discussion}
\label{sec:discussion}

Next, we highlight some interesting observations that we made during our experiments, which may need to be considered by an adversary carrying out the above de-anonymization attack. Further, we also list some additional adversarial optimizations that could be applied to the proposed framework and identify potential mitigation strategies against this threat. 

\begin{figure}[h]
	\centering
	\begin{subfigure}[b]{0.49\linewidth}
		\centering
		\includegraphics[width=0.9\linewidth]{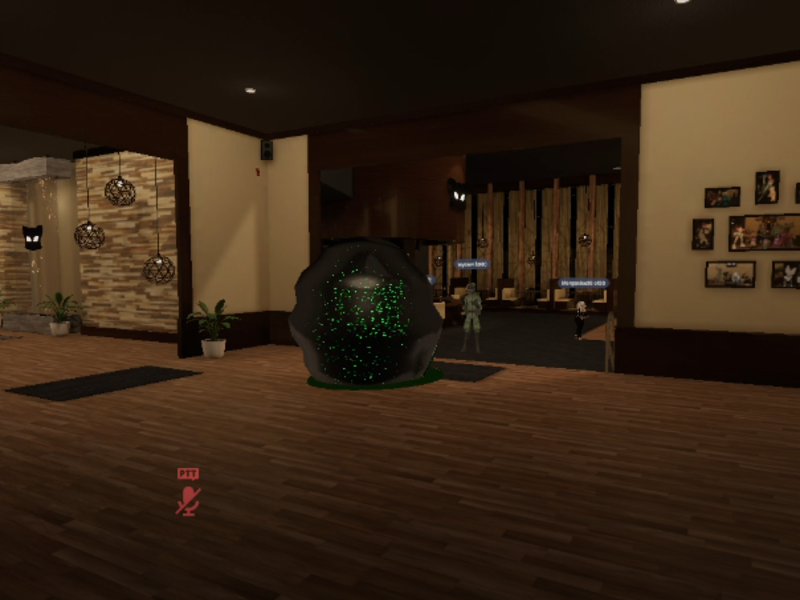}
		\caption{Meteoroid}
		\label{fig:vr-ball}
	\end{subfigure}
	\begin{subfigure}[b]{0.49\linewidth}
		\centering
		\includegraphics[width=0.9\linewidth]{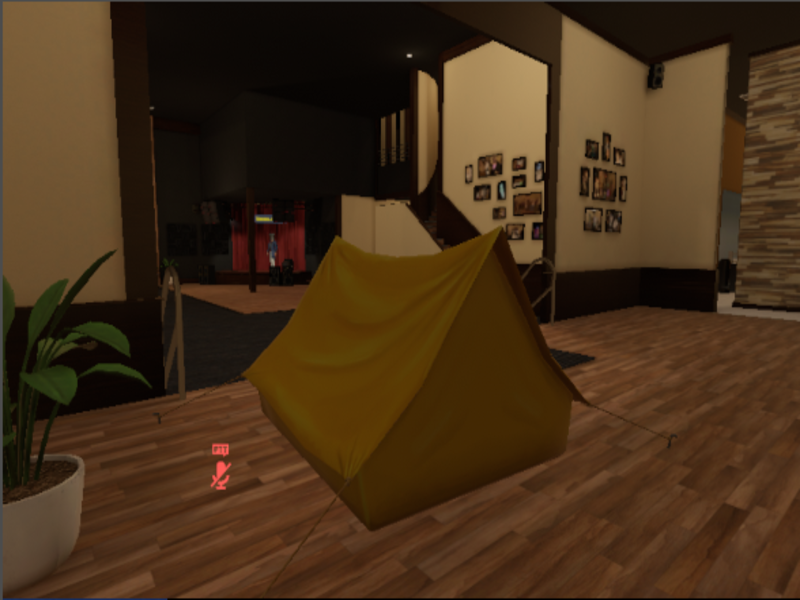}
		\caption{Tent}
		\label{fig:vr-tent}
	\end{subfigure}
	\caption{Examples of object spawning.}
	\label{fig:vr-objects}
\end{figure}

\noindent 
\textbf{Object Spawning.} 
During out experiments, we observed random objects, for example, a tent (\cref{fig:vr-tent}) and a meteoroid  (\cref{fig:vr-ball}), being spawned arbitrarily and at random locations within the VRChat worlds.
While the reason for these arbitrary objects appearing was unclear, depending on their location, they could interfere with the adversary's viewpoint by blocking his (visual) line-of-sight to the target.
In addition to randomly appearing stationary objects, we have also sometimes observed arbitrary appearances of moving non-playable characters which can also impact the adversary's view of the target.
In summary, an adversary should plan for such arbitrary obstructions during visual data collection, and perhaps employ multiple viewpoints (or perspectives) to the target user in order to overcome this issue, similar to what we do in our experiments.

\begin{figure}[h]
	\centering
	\begin{subfigure}[b]{0.49\linewidth}
		\centering
		\includegraphics[width=0.9\linewidth]{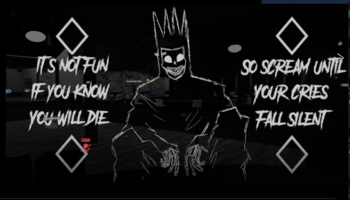}
		\caption{Injected image.}
		\label{fig:vr-threat-Img}
	\end{subfigure}
	\begin{subfigure}[b]{0.49\linewidth}
		\centering
		\includegraphics[width=0.9\linewidth]{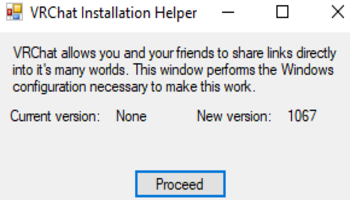}
		\caption{Injected link.}
		\label{fig:vr-install}
	\end{subfigure}
	\caption{Examples of image/link injection.}
	\label{fig:vr-threat}
\end{figure}

\noindent 
\textbf{Image \& Link Injections.} 
Another challenge an adversary could face in the virtual world (especially, public worlds) while collecting visual data (corresponding to the target) is random and uninitiated interactions with other VR users. During our experiments, we observed random users positioning themselves in front of our adversary (and its view), thus blocking his line-of-sight (to the target) and impacting the attack. While a mobile adversary may be able to adjust his position (within the virtual room) to regain view of the target, a stationary adversary may be unable to do it and thus unable to record useful visual data for the attack.  Other forms of interactions (by other users with our adversary) could include sharing of images and links, which could also disrupt the visual data recording by the adversary. For instance, during our experiments we observed that when an image is shared (see \cref{fig:vr-threat-Img}) by a VRChat user (with our adversary), it overlays a transparent image on top of the adversary's viewpoint, rendering the visual data collected by him ineffective during that period. Similarly, we also observed that sharing of links can also have undesirable effects on the adversary's avatar (\cref{fig:vr-install}), rendering it ineffective in collecting useful visual data.

\noindent 
\textbf{Detecting and Ousting Suspicious Avatars.} 
The VRChat service employs an anti-cheat software which attempts to detect bots, inactive avatars, and avatars who misuse VRChat terms of services and kicks them out or bans them from the service. %
During our experiments, we did observe that some of our adversarial avatars, especially stationary avatars, were kicked out of the room (being monitored) or even banned altogether from VRChat. Although the main reasons (could be the anti-cheat software or other users reporting our adversarial avatars) behind such kick-outs or bans are unclear to us, we believe this could present a significant obstacle to an adversary attempting to accomplish the proposed attack. 
In order to continue collecting visual data in the presence of such room kick-outs and bans, an adversary would need to find ways to circumvent such ``anti-cheat" measures or be ready to deploy backup avatars, similar to what we did during our experiments.

\noindent 
\textbf{Additional Optimization.} 
In addition to the optimizations we presented earlier in \cref{eval-scalability}, an adversary can carry out additional optimizations as part of the framework to improve the overall accuracy by further reducing the number of incorrect correlations. For example, suppose that an adversary has collected visual and motion data over multiple sessions/days. It is highly unlikely that a correlation between motion data of two (or more) unique people/users to a target avatar will repeat over a span of multiple independent observed virtual reality sessions. To utilize this factor, the adversary has to first increase the activity-based filtering threshold ($t$) for all the observed sessions/days. %
With a higher allowable mismatch between the activity sequences, the adversary is more likely to include the target user's identity in the rankings across all of the sessions. Thereafter, with elimination of identities not present in rankings of all the sessions, the combined ranking/search set will reduce drastically, increasing the probability of correct correlation.

\noindent 
\textbf{Active Mitigation Measures.} 
The best mitigation for the de-anonymization attack presented in this work is to fully decouple the visual and motion sensor data by not making the motion data available to the adversary when users are in virtual environments. This can be accomplished through various means such as increasing user awareness of such threats, not wearing/carrying smart mobile devices (with in-built motion sensors) while using VR services or through appropriate user-notifications at the beginning of VR sessions.
If the smart mobile device(s) is synced with the VR device, access to the mobile device motion sensor could also be automatically and appropriately regulated while the user is in a virtual reality session. 
Alternatively, another option to protect against such attacks would be to use non-humanoid avatars or a humanoid avatar with adversarial patches \cite{thys2019fooling}. 
Adversarial patches typically overlay an image patch on a target image object (in our case, an avatar), causing some pre-trained machine learning or deep learning classifier into misclassifying the object. An appropriate adversarial patch on the user's chosen avatar would prevent recognition of the humanoid character in our framework, thus preventing accurate generation of the activity-vector series required for correlation. %

\section{Conclusion}

We proposed a novel framework to correlate anonymous avatars in virtual worlds with identified out-of-band motion sensor data. Our work highlights a newfound privacy risk to users of the growing VR ecosystem. Specifically, VR users can be vulnerable to de-anonymization attack if they carry a smartphone or wear a smartwatch while using a VR system. Our evaluation of the proposed framework is a step towards demonstrating the feasibility of such an attack, utilizing real-world data from human participants. Through our empirical analyses, we were able to optimize framework parameters, improve scalability, and identified current limitations and potential for further improvements.

\bibliographystyle{ACM-Reference-Format}
\bibliography{main}

\appendix

\newpage

\section{Framework Details}

\begin{figure}[h]
	\centering
	\includegraphics[width=0.9\linewidth]{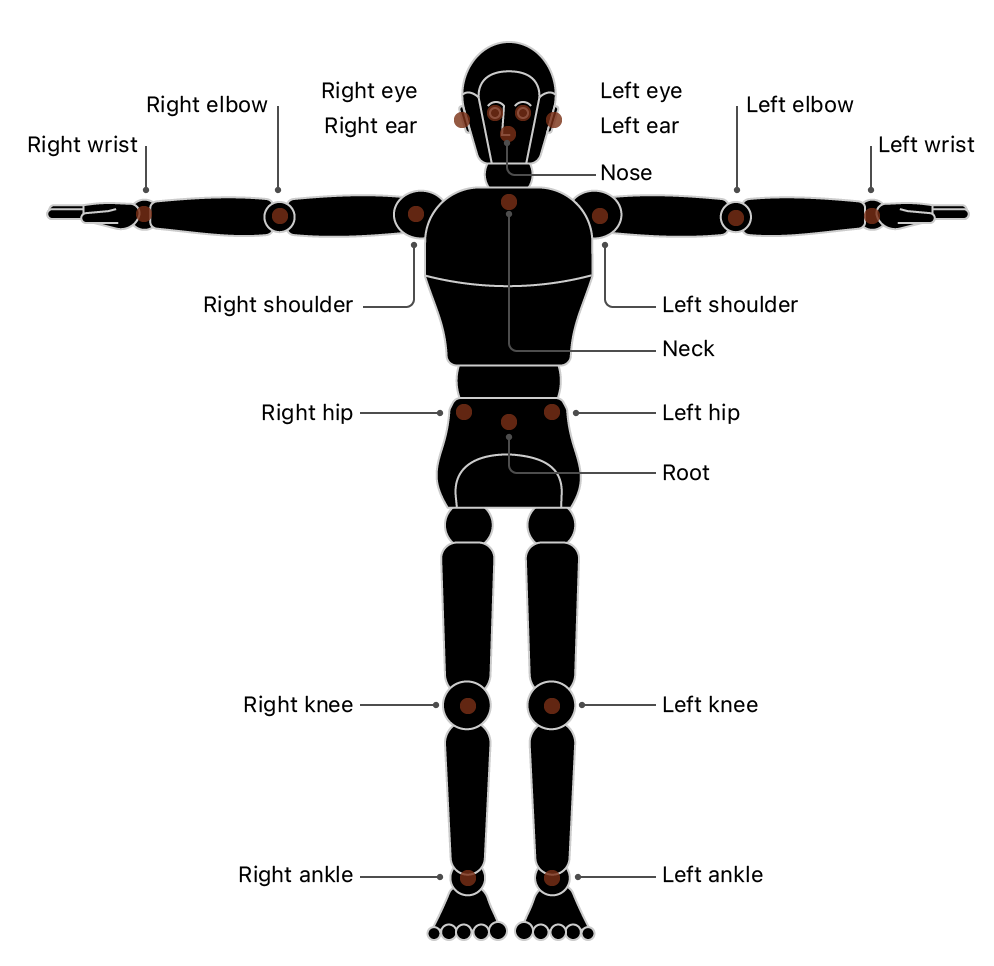}
	\caption{Keypoints on a human user or humanoid avatar.}
	\label{fig:keypoints}
\end{figure}

\begin{algorithm}[h]
	\small
	\caption{Correlation Algorithm.}
	\label{alg:keystroke-detection}
	\begin{algorithmic}[1]
		\State  \textbf{Input:}
		\State \hskip 1em  $video[]$  \Comment{Video's activity-vectors series}
		\State \hskip 1em  $ motion[]$ \Comment{Motion's activity-vectors series}
		\State \hskip 1em  $t$ \Comment{Filtering Threshold}
		\State \textbf{Output: } 
		\State \hskip 1em $ranked[]$ \Comment{Ranked list of correlated motion/video indexes with maximum Spearman's rank correlation coefficient}
		\Procedure{Correlate}{}%

		\State $correlated[]$ \Comment{Maps motion indexes to correlated  video indexes} %
		\State $unranked[]$ \Comment{Unranked list of correlated motion/video indexes with maximum Spearman's rank correlation coefficient}

		\For{ $i$ in range($video.size() - 1$)}
			\For{ $j$ in range($motion.size() - 1$)}
				\If{$HammingDistance(video[i], motion[j]) < t$ }
				\State $correlated[i].append(j)$
				\EndIf
			\EndFor
		\EndFor
		
		\For{ $i$ in range($video.size() - 1$)}
			\For{$j$ in range($correlated[i].size() - 1$)}
			\State $m_{idx} = correlated[i][j]$  \Comment{motion index}
			\State $maxSpearman = max(Spearman(video[i], motion[m_{idx}]))$
			\State $unranked[i].append(\{maxSpearman, m_{idx}\})$
			\EndFor
			\State $ranked[i] = unranked[i].sort() $ 
			\Comment{ sorted based on Spearman’s rank correlation coefficient}
		
		\EndFor
		\EndProcedure
	\end{algorithmic}
\end{algorithm}

\onecolumn
\section{Controlled Activity Sets in Data Collection}
\begin{table}[H]
    \centering
    \small
    \caption{List of controlled actions performed by participants in the real and virtual reality worlds.}
      \begin{tabular}{rr}
      \toprule
      \multicolumn{1}{c}{\textbf{Action Types}} & \multicolumn{1}{c}{\textbf{Action Description}} \\
      \midrule
      \multicolumn{1}{l}{Head-based} & \multicolumn{1}{l}{Looking [left, right, up, down]} \\
      \multicolumn{1}{l}{} & \multicolumn{1}{l}{Rotating the head in [clockwise, anti-clockwise] directions} \\
      \multicolumn{1}{l}{Arm-based} & \multicolumn{1}{l}{Raising [left, right, both] arms in [forward, upward, sideward] directions} \\
      \multicolumn{1}{l}{} & \multicolumn{1}{l}{Rotating [left, right, both] arms in [clockwise, anti-clockwise] directions} \\
      \multicolumn{1}{l}{} & \multicolumn{1}{l}{Stretching arms [forward, upward, sideward]} \\
      \multicolumn{1}{l}{Palm-based} & \multicolumn{1}{l}{Handshaking with [left, right, both] arms} \\
      \multicolumn{1}{l}{} & \multicolumn{1}{l}{Waving with [left, right, both] arms in [forward, upward] directions} \\
      \multicolumn{1}{l}{} & \multicolumn{1}{l}{Thumbs up and down with [left, right, both] arms forward} \\
      \multicolumn{1}{l}{} & \multicolumn{1}{l}{Clapping with hands forward} \\
      \multicolumn{1}{l}{Leg-based} & \multicolumn{1}{l}{Stepping along [left, right, forward, backward] directions} \\
      \multicolumn{1}{l}{} & \multicolumn{1}{l}{Walking diagonally towards [left, right, forward, backward] directions} \\
      \multicolumn{1}{l}{} & \multicolumn{1}{l}{Raising [left, right] knee} \\
      \multicolumn{1}{l}{Combination-based} & \multicolumn{1}{l}{[Twisting hip, turning body around] in [clockwise, anti-clockwise] directions} \\
      \multicolumn{1}{c}{} & \multicolumn{1}{l}{Crouching or squatting, Jumping up and down} \\
      \multicolumn{1}{c}{} & \multicolumn{1}{l}{Sitting on the [floor, chair]} \\
      \multicolumn{1}{c}{} & \multicolumn{1}{l}{Exploring [public, private] instances} \\
      \multicolumn{1}{c}{} & \multicolumn{1}{l}{[Walking, running] in [straight, zig-zag] paths} \\
      \multicolumn{1}{c}{} & \multicolumn{1}{l}{[Talking, browsing] smartphone in [portrait, landscape] modes} \\
      \multicolumn{1}{c}{} & \multicolumn{1}{l}{Fiddling with an object} \\
      \multicolumn{1}{c}{} & \multicolumn{1}{l}{Picking up objects placed on the [floor, table]} \\
      \bottomrule
      \end{tabular}%
    \label{tab:activities}%
\end{table}

\twocolumn

\end{document}